\documentclass[12pt,american]{article}
\usepackage{lmodern}

\usepackage[T1]{fontenc}
\usepackage[latin9]{inputenc}
\usepackage{geometry}
\usepackage{color}
\usepackage{verbatim}
\usepackage{amsmath}
\usepackage{graphicx}
\usepackage{setspace}
\usepackage{esint}
\usepackage{babel}
\usepackage{soul}
\usepackage{xcolor}
\usepackage{breqn}
\usepackage{hhline}
\usepackage{rotating}
\usepackage{booktabs}
\usepackage{amsfonts}
\usepackage{mathbbol}
\usepackage[normalem]{ulem}
\usepackage[longnamesfirst]{natbib}

\definecolor{orange}{RGB}{255,127,0}
\usepackage{tabulary}
\newcolumntype{L}{>{\centering\arraybackslash}m{8cm}}
\usepackage{mathrsfs, amsmath}
\usepackage{graphicx}
\usepackage{siunitx} 
\usepackage{placeins}
\usepackage{array}
\usepackage{multirow}
\usepackage{bbm}
\usepackage{booktabs}
\usepackage{subcaption}
\usepackage[flushleft]{threeparttable}
\usepackage{hyperref}
\usepackage{transparent}
\usepackage{geometry,mathtools}
\usepackage{xcolor}
\usepackage{changepage}
\usepackage{siunitx}
\usepackage{anyfontsize}
\usepackage{colortbl} 
\usepackage{cleveref}
\usepackage{comment}
\usepackage{soul}

\DeclareMathOperator*{\argmax}{arg\,max}

\newcommand{\xuparrow}[1]{%
  {\left\uparrow\vbox to #1{}\right.\kern-\nulldelimiterspace}
}

\newcolumntype{g}{>{\color{lightgray}}c}

\makeatother
\makeatletter

\usepackage{bigints}
\usepackage{babel}
\usepackage{soul}
\usepackage{xcolor}
\makeatother

\begin{document} 

\title{Macroeconomics of Racial Disparities: Discrimination, Labor Market, and Wealth}

\author{Srinivasan Murali%
\thanks{srinivasan.murali@flame.edu.in. 
Department of Economics, FLAME University, Pune, Maharashtra 412115, India.%
} ~~~~ Guanyi Yang%
\thanks{Corresponding author: gyang@ColoradoCollege.edu. Department of Economics \& Business, Colorado College, Palmer Hall, 14 E. Cache la Poudre St., Colorado Springs, CO 80903, USA.%
}\\}

\date{January 2026}

\maketitle

\begin{abstract}
\begin{onehalfspace}
This paper examines the impact of racial discrimination in hiring on employment, wages, and wealth disparities between black and white workers. Using a labor search-and-matching model with racially prejudiced and non-prejudiced firms, we show that labor market frictions sustain discriminatory practices as an equilibrium outcome. These practices account for 57\% of the racial unemployment gap, 48\% of the average wage gap, and 16\% of the median wealth gap. Discriminatory hiring also increases unemployment and wage volatility for black workers, increasing their labor market risks over the business cycle. Eliminating prejudiced firms reduces these disparities and improves the welfare of black workers as well as the overall economic welfare.\\
{\color{white}-}\\
\noindent\textit{JEL classification: D14, E21, J15, J64, J65} \\
\textit{Keywords:} search-and-matching, heterogeneous agents, hiring discrimination, racial inequality, labor market frictions, unemployment, wage gaps, wealth distribution, business cycles \end{onehalfspace}

\end{abstract}
\addtocounter{page}{-1}
\thispagestyle{empty}
\newpage{}

\section{Introduction}

Market competition and increased access to information are expected to eliminate discrimination over time. Nonetheless, substantial empirical evidence shows that discriminatory practices persist across markets, perpetuating significant racial disparities.\footnote{For example, \citet[][]{couch2010last, biddle2013wage, kuhn2020income, derenoncourt2021minimum, lippens2023state, lang2020race}. Resume studies consistently reveal racial discrimination even when controlling for candidate qualifications \citep[e.g.,][]{bertrand2004emily, bertrand2017field, quillian2017meta, kline2022systemic}. Gaps in labor income and wealth between black and white households have persisted long after the Civil Rights Movement \citep[e.g.,][]{cajner2017racial, derenoncourt2023changes}. \citet{small2020sociological} argues that discrimination reinforces itself across domains and constitutes a form of market failure.} Only recently has macroeconomic research begun exploring racial disparities.\footnote{See, for example, \citet{nakajima2021monetary}, \citet{aliprantis2023dynamics}, \citet{boerma2021reparations}, \citet{ganong2020wealth}, \citet{lee2021minority}.} Yet, the role of discriminatory practices in shaping these disparities remains underexplored. Developing a general equilibrium framework to explain the persistence of discrimination and its consequences is crucial for understanding the macroeconomic impacts of racial disparities and informing effective policymaking.

This paper constructs a search-and-matching model to examine the macroeconomic effects of persistent racial discrimination in hiring. We find that hiring discrimination explains roughly 57\% of the black-white unemployment rate gap, 48\% of the average wage gap, and 16\% of the median wealth gap in the steady state. Moreover, discriminatory hiring amplifies the volatility of unemployment and wages among black workers over the business cycle. Eliminating hiring discrimination not only improves the welfare of black workers but also enhances overall economic welfare.

The model distinguishes between two types of firms: prejudiced firms, which hire only white workers, and non-prejudiced firms, which hire without regard to race. This structure implies that black workers can find employment only at non-prejudiced firms, whereas white workers may be hired by either type. As is standard in search-and-matching frameworks, firms are synonymous with jobs or vacancies. We interpret prejudiced firms as vacancies associated with prejudiced hiring decision-makers who exclude black applicants. Modeling discrimination at the vacancy level, therefore, captures differential access to job opportunities rather than differences in worker productivity or firm technology.

This interpretation aligns closely with extensive empirical evidence on racial disparities in hiring. Audit studies show that otherwise identical resumes receive about 50\% more callbacks when associated with white-sounding names \citep{bertrand2004emily}, and a comprehensive meta-analysis finds that white applicants receive, on average, 36\% more callbacks across settings \citep{quillian2017meta}. While discrimination may also arise in wage-setting, promotions, or separations, we focus on hiring discrimination because it operates at the point of entry into employment. From a macroeconomic perspective, focusing on hiring directly affects job-finding probabilities and allows us to isolate the aggregate and distributional consequences of discrimination in a general equilibrium setting, without introducing additional channels that would confound its distinct macroeconomic effects.

In the model, matched worker-firm pairs bargain over wages to maximize joint surplus. When calibrated, the model endogenously generates lower job-finding rates, higher unemployment, and lower wages for black workers. Even in the absence of additional financial frictions, these labor market differences translate into sizable racial gaps in wealth accumulation.

The main message of this paper is that labor market frictions can sustain hiring discrimination as an equilibrium outcome. Our calibration shows that prejudiced firms incur higher vacancy posting costs to discriminate against black workers. Yet, they sustain positive profits by maintaining a lower turnover rate upon hiring white workers. Moreover, hiring discrimination affects the equilibrium economy through firms' vacancy postings. Prejudiced firms compete with non-prejudiced firms for white workers, driving up their wage rates across the economy. This wage pressure reduces the expected profit of non-prejudiced firms, leading them to post fewer vacancies overall. The resulting decline in job openings at non-prejudiced firms directly limits employment opportunities for black workers, contributing to lower wages, higher unemployment, and adverse wealth accumulation. In a counterfactual economy without prejudiced firms, these racial disparities narrow substantially.
 
Eliminating discriminatory hiring also yields significant welfare gains. Black workers experience a 12.9\% increase in welfare as non-prejudiced firms expand their vacancy postings, which in turn improves their job opportunities and wages. Although white workers incur a modest 0.5\% welfare decline due to the loss of exclusive job opportunities, the overall welfare of the economy rises by 1.5\% because the gains for black workers more than offset the losses for white workers.

We further incorporate aggregate shocks into our benchmark economy to examine how hiring discrimination shapes business cycle dynamics. Our model replicates key empirical patterns: a countercyclical and volatile racial unemployment gap, as well as a more procyclical and volatile wage response among black workers. In a counterfactual model without prejudiced firms, the volatility of unemployment rates and wages for black workers is substantially reduced relative to white workers. Moreover, hiring discrimination amplifies the volatility of racial disparities in consumption and wealth over the business cycle. These findings suggest that firm-side hiring discrimination plays a crucial role in driving the distinct business cycle dynamics of black and white workers, which cannot be fully explained by demographic factors alone \citep{cajner2017racial, boulware2019labor, boulware2024explains}.

This paper contributes to the ongoing discussion on the persistence of racial inequality by focusing on the disparate labor market conditions of black and white workers. Numerous studies have documented racial discrimination in pay and employment opportunities that persist despite policy interventions \citep[e.g.,][]{coate1993will, black1995discrimination, rosen1997equilibrium, bertrand2004emily, manduca2018income}. Recent empirical research stresses that racial wage and employment gaps persist across education levels, skill sets, and cohorts. These disparities are not solely attributable to differences in individual qualifications \citep[e.g.,][]{pena2018skills, cheng2019educational}. Building on this insight, our paper develops a model in which discriminatory hiring practices endogenously perpetuate labor market inequalities. Our work adds to the literature by providing a theoretical framework that explains how hiring discrimination can be sustained as an equilibrium outcome in the labor market.

While much of the literature focuses on static disparities, fewer studies document how macroeconomic fluctuations impact black and white workers differently. Notably, \citet{couch2010last} demonstrate that black workers are disproportionately affected during recessions, being the last hired in upturns and the first fired in downturns. \citet{biddle2013wage} find that the discriminatory wage gap is procyclical, widening during economic expansions. Similarly, \citet{cajner2017racial} show that black workers experience higher unemployment rate volatility and are more likely to be involuntarily employed part-time. \citet{daly2020labor} suggest that limited employment opportunities for black workers contribute to the growing racial earnings gap. Recent research by \citet{boulware2019labor, boulware2024explains} further highlights that racial disparities intensify during downturns, driven by increased hiring discrimination and differential sensitivity of black employment to cyclical fluctuations, effects that cannot be fully attributed to black workers' concentration in cyclically sensitive industries. Our paper adds to this body of work by evaluating the impact of hiring discrimination on the cyclical aspects of racial disparities, examining how discriminatory practices exacerbate the volatility and cyclicality of employment and wages for black workers.

An emerging strand of literature explores racial disparities in wealth accumulation. Studies such as \citet{derenoncourt2023changes}, \citet{derenoncourt2022wealth}, \citet{kuhn2020income}, \citet{barsky2002accounting}, and \citet{mcintosh2020examining} document significant wealth gaps between black and white households. For example, \citet{derenoncourt2023changes} provides a historical account of wealth segregation over the past 150 years, showing persistent and substantial disparities. \citet{boerma2021reparations} and \citet{aliprantis2023dynamics} analyze the impact of historical discrimination on earnings, bequests, and capital returns within steady-state models that do not consider aggregate risks. Furthermore, \citet{ganong2020wealth} demonstrates that income risks affect individuals differently across racial groups due to wealth differences. Building on these findings, \citet{bartscher2021monetary} and \citet{lee2021minority} discuss how monetary policy can have disparate effects on workers of different racial backgrounds. Our paper contributes to this literature by showing that employer discrimination not only affects immediate labor market outcomes but also spills over to long-term wealth accumulation disparities between black and white workers.

Germane to our project, \citet{nakajima2021monetary} develops a search-and-matching model to examine how monetary policy exacerbates racial differences in the labor market. Our model explicitly distinguishes between prejudiced and non-prejudiced firms in the hiring process and directly examines how hiring discrimination impacts black workers and the broader economy. We analyze the spillover effects of discriminatory hiring practices on both labor and wealth disparities, offering a comprehensive theoretical understanding of these dynamics over the long run and across business cycles.

In a broader context, our paper joins the growing discussion on the distributional impact of economic growth and macroeconomic policies by focusing on the heterogeneous outcomes for black and white workers. This aligns with studies like \citet{caballero1994cleansing}, \citet{jaimovich2020job}, and \citet{heathcote2020}, which provide evidence that recessions disproportionately hurt disadvantaged individuals. \citet{borella2018aggregate} shows that incorporating gender differences in a life-cycle model improves its empirical fit, emphasizing the importance of accounting for demographic heterogeneity. \citet{krusell1998income} concludes that heterogeneity in wealth does not alter business cycle fluctuations. Yet, \citet{yum2023heterogeneity} shows that heterogeneity could generate large aggregate fluctuations when introducing non-convexity in budget constraints through progressive tax. Our paper demonstrates that incorporating racial differences into labor market models enhances their ability to capture business cycle fluctuations. It also provides insights into how racial wedges in labor search processes transmit individual risks and heterogeneity into aggregate economic dynamics.

The rest of the paper proceeds as follows. \Cref{sec: model} lays out the theoretical framework. 
\Cref{sec: cali} discusses the calibration strategy. \Cref{sec: steady} examines the steady-state implications of racial discrimination. \Cref{sec: biz} discusses the business cycle implications. 
\Cref{sec: conclude} concludes the paper.

\section{Model}\label{sec: model}

This section presents a model of labor market discrimination without aggregate uncertainty. We use this framework to understand the role of hiring discrimination in driving the racial gaps in labor market outcomes and wealth.

\subsection{Environment}

The economy consists of a unit measure of workers, who are either black or white, \( R \in \{bl, wh\} \), and firms that post vacancies subject to search and matching frictions. Firms differ in their hiring behavior. Prejudiced firms (\( p \)) exclude black applicants, while non-prejudiced firms (\( np \)) hire workers without regard to race. A match between a worker and a firm corresponds to a filled job, so firms and jobs are treated as equivalent objects in the model. As a result, black workers can receive job offers only from non-prejudiced firms, whereas white workers may receive offers from both types. To simplify notation, we suppress time subscripts and use a prime (\('\)) to denote next-period values.

Firms are modeled as vacancies that differ only in their hiring rules, rather than in their production technologies. Prejudiced firms are thus represented as vacancies that are inaccessible to black workers by assumption. This reduced-form approach isolates differences in access to job opportunities while remaining agnostic about the underlying source of discrimination. The modeling choice is motivated by extensive evidence of racial disparities in hiring. Audit studies show that otherwise identical resumes receive fewer callbacks when associated with black-sounding names \citep{bertrand2004emily}, a pattern that has been documented across contexts in a comprehensive meta-analysis \citep{quillian2017meta}.

Workers are either employed (with $p$ or $np$ firm) or unemployed. Those who become unemployed in the current period receive unemployment benefits and continue to receive them in the future with probability $P_e$. The workers face idiosyncratic productivity shocks $s$, following an AR(1) process $\log(s') = \rho_s \log(s) + \epsilon_s$, with $\epsilon_{s} \stackrel{iid}{\sim} N(0,\sigma_{s}^2)$.\footnote{To ease exposition, we drop the time subscripts and use a prime symbol (') to denote variables in the next period.} Workers also differ in their asset holdings. They can save using risk-free assets to partially insure themselves against labor market risks. Following \cite{mukoyama2013understanding}, workers also receive race-specific extreme wealth shocks with probability $\epsilon_{R} \in \{\epsilon_{bl}, \epsilon_{wh}\}$. Upon realization of the shock, a worker loses all their wealth. Altogether, workers are heterogeneous across race ($R$), labor market status ($e$), idiosyncratic productivity ($s$), and wealth ($a$). The endogenous distribution of workers is summarized by $\mu(e,R,s,a)$. In the steady state, individual workers may move around the distribution, but the overall distribution remains stationary. Thus, $\mu$ is not a state variable in the steady state model but becomes an aggregate state variable in the augmented model with aggregate shocks discussed in \Cref{sec: biz}.

\subsection{Labor market search and matching}

The total number of unemployed workers, denoted as \( u \), is the sum of unemployed black workers \( (u_{bl}) \) and unemployed white workers \( (u_{wh}) \). There are \( v_{np} \) vacancies available in the non-prejudiced market, while the number of vacancies in the prejudiced market is \( v_p \). Non-prejudiced firms search among both black and white unemployed workers, resulting in a non-prejudiced market tightness defined as \( \theta_{np} = \frac{v_{np}}{u} \). In contrast, prejudiced firms only hire unemployed white workers, which gives the prejudiced market tightness \( \theta_{p} = \frac{v_p}{u_{wh}} \). Following the works of \citet{den2000job} and \citet{petrosky2018endogenous}, unemployed workers and vacant firms match through a constant returns to scale matching function
\begin{equation}
    M(u,v) = \frac{uv}{(u^{\iota}+v^{\iota})^{1/\iota}}
\end{equation}
with $\iota>0$. As documented by \citet{den2000job}, this functional form ensures that the matching probabilities stay within 0 and 1. The probability for an unemployed worker to match with a vacant $np$ firm is $f(\theta_{np}) =M(u,v_{np})/u = (1+\theta_{np}^{-\iota})^{-1/\iota}$, while the probability that a white unemployed worker matches with a vacant $p$ firm is $f(\theta_{p}) =M(u_{wh},v_{p})/u_{wh} = (1+\theta_{p}^{-\iota})^{-1/\iota}$. Correspondingly, the probability of filling a vacant $np$ firm is $q(\theta_{np}) = M(u,v_{np})/v_{np} = (1+\theta_{np}^{\iota})^{-1/\iota}$, while the probability of filling a vacant $p$ firm is $q(\theta_{p}) =M(u_{wh},v_{p})/v_{p} = (1+\theta_{p}^{\iota})^{-1/\iota}$. In addition, non-prejudiced matches are destroyed with a probability of $\lambda_{np}$, while prejudiced matches separate at the rate of $\lambda_{p}$.

This setup is equivalent to an environment in which all workers search in a unified labor market, but prejudiced firms reject black applicants upon meeting them. Rather than modeling these unsuccessful encounters explicitly, we adopt an equivalent formulation in which black workers direct their search toward non-prejudiced firms, while white workers search across all vacancies. This representation simplifies the analysis while preserving differential job finding opportunities across racial groups. Empirically, persistent and cross-sectional differences in race-specific job finding and separation rates account for a substantial share of observed differences in unemployment dynamics between black and white workers \citep{boulware2024explains}. In our framework, because prejudiced firms hire only white workers and non-prejudiced firms are race-blind, job finding and separation probabilities are parameterized at the firm type level to generate these observed racial differences.

\subsection{Unemployment Insurance}

Unemployment insurance is characterized by the replacement rate $h$, eligibility probability $P_e$, and maximum coverage level $\chi$. Following \cite{setty2021provision}, eligible workers receive unemployment benefits $b(R,s,a) = min\{h\bar{w}(R,s,a),\chi\}$, where $\bar{w}(R,s,a)$ is the counterfactual wage earned by an employed worker with race $R$, productivity $s$, and wealth $a$. We adopt the counterfactual wage to ease the computation burden of tracking wage history. Similar to \cite{mitman2015optimal}, newly unemployed workers receive unemployment benefits with certainty and continue to receive benefits next period with probability $P_e$. If an unemployed worker loses their eligibility to receive benefits, they continue to remain ineligible in the future. Unemployment benefits are funded through a proportional tax $\tau$ on the labor income, and the government sets $\tau$ to balance its budget.\footnote{We intentionally model a more realistic and complex unemployment insurance structure to capture the racial disparities in the incidence and take-up of unemployment insurance. This helps us generate racial differences in income and wealth distribution, given the asymmetric labor market and wealth risks.}
 
\subsection{Workers}

The value function of an employed individual with race $R$, productivity $s$, asset $a$, and working with an $np$ firm is given by $W_{np}(R,s,a)$, while that of a white worker employed with a $p$ firm is given by $W_p(wh,s,a)$. Since the unemployment benefit is indexed to the worker's counterfactual wage, the values of the unemployed workers eligible for benefits depend on whether they worked with a non-prejudiced (with value function $U^I_{np}(R,s,a)$) or a prejudiced (with value function $U^I_p(wh,s,a)$) firm previously. An unemployed worker who is not eligible for unemployment benefits has a value of $U^N(R,s,a)$ over their lifetime. All the workers discount their future utility by $\beta$. Similar to \cite{nakajima2012business} and \cite{setty2021provision}, we assume that workers cannot borrow. This imposes an exogenous constraint of $a'\geq 0$ on all workers.

\subsubsection{Employed with \textit{np} firm}

\Cref{eqn:worker1} describes an employed worker of race $R$ with productivity $s$ and asset holdings $a$, working in an $np$ firm. The worker chooses consumption $c$ and future savings $a'$ to maximize their discounted lifetime utility. Their income consists of period wage, $\omega_{np}$, net of payroll tax $\tau$, current savings $(1+r)a$, and dividends $d$. The expectation of the worker's future value is taken over the idiosyncratic productivity shock $s$ and the race-specific probability of the extreme wealth shock $\epsilon_R$. 

\begin{equation}
\begin{aligned}
         W_{np}(R,s,a) &= \max_{c,a' \geq 0} \Big\{ u(c) + \beta  \sum_{s'} \pi_{ss'}\big[ (1-\epsilon_R) \hat{W}_{np}(R,s',a') + \epsilon_R \hat{W}_{np}(R,s',0)\big]\Big\}\\ 
        \text{ s.t. }\\
         c + a' &= (1-\tau)\omega_{np}(R,s,a)  + (1+r)a + d
\end{aligned}
\label{eqn:worker1}
\end{equation}

The worker receives a job destruction shock specific to $np$ firms and becomes unemployed with probability $\lambda_{np}$. Otherwise, they continue to remain employed with probability $1 - \lambda_{np}$.  If unemployed, the worker receives unemployment benefits and realizes a value of $U^I_{np}$. Thus, the future value of an employed worker conditional on the realization of wealth shock is given by \Cref{eqn:worker1.2}.

\begin{equation}
    \hat{W}_{np}(R,s',a') = \lambda_{np} U^I_{np}(R,s',a') + (1-\lambda_{np})W_{np}(R,s',a') 
\label{eqn:worker1.2}    
\end{equation}

\subsubsection{Employed with \textit{p} firm}

\Cref{eqn:worker2} describes the case of a worker employed by a $p$ firm. It is analogous to the previous case, but we only consider white workers since prejudiced firms do not hire black workers.

\begin{equation}
\begin{aligned}
        W_{p}(wh,s,a) &= \max_{c,a' \geq 0} \Big\{ u(c) + \beta \sum_{s'} \pi_{ss'} \big[ (1-\epsilon_{wh}) \hat{W}_{p}(wh,s',a') + \epsilon_{wh} \hat{W}_{p}(wh,s',0)   \big]\Big\}\\ 
        \text{ s.t. }\\
        c + a' &= (1-\tau)\omega_{p}(wh,s,a)  + (1+r)a + d 
\end{aligned}
\label{eqn:worker2}
\end{equation}
The matched worker receives a job destruction shock specific to $p$ firms, with probability $\lambda_{p}$. A worker losing the job in the current period is eligible for unemployment benefits and earns value $U^I_{p}$ in the next period. Conditional on realizing the wealth shock, an employed worker's future value expands to \Cref{eqn:worker2.2}.

\begin{equation}
    \hat{W}_{p}(wh,s',a') = \lambda_{p} U^I_{p}(wh,s',a') + (1-\lambda_{p})W_{p}(wh,s',a')
\label{eqn:worker2.2}    
\end{equation}

\subsubsection{Unemployed and eligible workers}

Since unemployment insurance is proportional to the counterfactual wage, the value obtained by an unemployed and eligible worker depends on whether the past employment was in a $p$ or in an $np$ firm. In addition, unemployed white workers can receive job offers from both $p$ and $np$ firms, while black workers can get matched only with $np$ firms. 

\quad\\  
\noindent \textbf{Unemployed black worker from $np$ firm} 

\Cref{eqn:worker3} describes the value of an unemployed black worker eligible for benefits. The worker encounters the extreme wealth shock with probability $\epsilon_{bl}$. The worker receives unemployment insurance payout, $b_{np}(bl,s, a)$, net of labor income tax rate $\tau$, previous savings, and lump-sum dividend transfer $d$. 

\begin{equation}
\begin{aligned}
        U^I_{np}(bl,s,a) &= \max_{c,a' \geq 0} \Big\{ u(c) + \beta \sum_{s'} \pi_{ss'}\big[ (1-\epsilon_{bl}) \hat{U}^I_{np}(bl,s',a') + \epsilon_{bl} \hat{U}^I_{np}(bl,s',0)   \big]\Big\}\\ 
             &\text{ s.t. }\\
        c + a' &= (1-\tau)b_{np}(bl,s,a) + (1+r)a + d
\end{aligned}
\label{eqn:worker3}
\end{equation}

Their future value $\hat{U}^I_{np}(bl,s',a')$ expands to \Cref{eqn:worker3.2}.
\begin{equation}
    \hat{U}^I_{np}(bl,s',a') = f(\theta_{np}) W_{np}(bl,s',a') + (1-f(\theta_{np})) [P_e U^I_{np}(bl,s',a') + (1-P_e)U^N(bl,s',a') ]  
\label{eqn:worker3.2}    
\end{equation}

The unemployed black worker finds a job with probability $f(\theta_{np})$, which moves them to a future value of $W_{np}(bl,s',a')$. Otherwise, they remain unemployed and continue to receive unemployment benefits with probability $P_e$ or lose their benefit with probability $(1-P_e)$. If one loses the benefit, the worker receives a value of $U^N(bl,s',a')$.

\quad\\  
\noindent \textbf{Unemployed white worker from $np$ firm}

Unemployed white workers are recruited by both $p$ and $np$ firms, and their value function is given by \Cref{eqn:worker4}. Like a black worker, a white worker encounters an extreme wealth shock with probability $\epsilon_{wh}$. They also have a budget constraint depending on their post-tax unemployment insurance payout, current savings, and a lump-sum dividend transfer. 

\begin{equation}
\begin{aligned}
        U^I_{np}(wh,s,a) &= \max_{c,a' \geq 0} \Big\{ u(c) +  \beta \sum_{s'} \pi_{ss'}\big[ (1-\epsilon_{wh}) \hat{U}^I_{np}(wh,s',a') + \epsilon_{wh} \hat{U}^I_{np}(wh,s',0)   \big] \Big\}\\
             &\text{ s.t. }\\
        c + a' &= (1-\tau)b_{np}(wh,s,a)  + (1+r)a + d.
\end{aligned}
\label{eqn:worker4}
\end{equation}

\Cref{eqn:worker4.2} describes the future value of an unemployed white worker conditional on the wealth shock. With probability $(1-f(\theta_{np}))f(\theta_p)$, the worker receives a job offer only from a $p$ firm, while they receive an offer only from a $np$ firm with probability $f(\theta_{np})(1-f(\theta_p))$. With probability $f(\theta_{np})f(\theta_{p})$, they receive offers from both $np$ and $p$ firms simultaneously. If the worker receives both $p$ and $np$ offers, they will choose the job with a higher lifetime value. The white worker will remain unemployed if they don't receive any offer (with probability $(1-f(\theta_p))(1-f(\theta_{np}))$). In this case, the worker continues to receive unemployment benefits with probability $P_e$, and lose their eligibility with probability $(1-P_e)$.

\begin{equation}
\begin{aligned}
        \hat{U}^I_{np}(wh,s',a') &= (1-f(\theta_{np}))f(\theta_p)W_p(wh,s',a') + f(\theta_{np})(1-f(\theta_p)) W_{np}(wh,s',a') \\
          &+ f(\theta_{np})f(\theta_{p})\max\{W_p(wh,s',a'),W_{np}(wh,s',a')\} \\      
        &+ (1-f(\theta_p))(1-f(\theta_{np})) \Big[ P_e U^I_{np}(wh,s',a') + (1-P_e)U^N(wh,s',a') \Big] 
\end{aligned} 
\label{eqn:worker4.2}
\end{equation}

\quad\\
\noindent \textbf{Unemployed white worker from $p$ firm}

\Cref{eqn:worker5} describes the problem faced by a white worker who last worked with a $p$ firm. The problem is identical to the previous case, except that they receive an insurance payout of $b_{p}(wh,s,a)$.

\begin{equation}
\begin{aligned}
        U^I_{p}(wh,s,a) &= \max_{c,a' \geq 0} \Big\{ u(c) +  \beta \sum_{s'} \pi_{ss'} \big[ (1-\epsilon_{wh}) \hat{U}^I_{p}(wh,s',a') + \epsilon_{wh} \hat{U}^I_{p}(wh,s',0)   \big] \Big\}\\
        &\text{ s.t. }\\
        c + a' &= (1-\tau)b_{p}(wh,s,a)  + (1+r)a + d
\end{aligned}
\label{eqn:worker5}
\end{equation}
Analogous to the previous case, the unemployed worker's future value conditional on the wealth shock is given by \Cref{eqn:worker5.2}. 
\begin{equation}
\begin{aligned}
        \hat{U}^I_{p}(wh,s',a') &= (1-f(\theta_{np}))f(\theta_p)W_p(wh,s',a') +  f(\theta_{np})(1-f(\theta_p)) W_{np}(wh,s',a') \\
         &+ f(\theta_{np})f(\theta_{p})\max\{W_p(wh,s',a'),W_{np}(wh,s',a')\} \\ 
        &+  (1-f(\theta_p))(1-f(\theta_{np}))  \Big[ P_e U^I_{p}(wh,s',a')  + 
         (1-P_e)U^N(wh,s',a') \Big]
\end{aligned}
\label{eqn:worker5.2}
\end{equation}

\subsubsection{Unemployed and ineligible workers}

Ineligible workers are those who have lost their eligibility for receiving unemployment insurance and are still searching for jobs. Once ineligible, they no longer become eligible again. Therefore, their consumption relies solely on their savings and lump sum dividends, regardless of their prior employment status.

\quad\\
\noindent \textbf{Black worker}

\Cref{eqn:worker6} describes the problem of an unemployed black worker searching for jobs without unemployment insurance. 

\begin{equation}
\begin{aligned}
        U^N(bl,s,a) &= \max_{c,a' \geq 0} \Big\{ u(c) + \beta \sum_{s'} \pi_{ss'}\Big[(1-\epsilon_{bl}) \hat{U}^N(bl,s',a') + \epsilon_{bl} \hat{U}^N(bl,s',0) \Big]\Big\}\\ 
            &\text{ s.t. }\\
        c + a' &= (1+r)a + d
\end{aligned}
\label{eqn:worker6}
\end{equation}
\Cref{eqn:worker6.2} expands the conditional future value of the black worker. The worker finds a job with an $np$ firm with probability \( f(\theta_{np}) \), and remains unemployed and ineligible otherwise.

\begin{equation}
\begin{aligned}
        \hat{U}^N(bl,s',a') &= f(\theta_{np}) W_{np}(bl,s',a') + (1-f(\theta_{np})) U^N(bl,s',a') 
\end{aligned}
\label{eqn:worker6.2}
\end{equation}

\noindent \textbf{White worker}

Similarly, \Cref{eqn:worker7} describes the problem of a white worker searching for jobs without unemployment insurance. 

\begin{equation}
\begin{aligned}
        U^N(wh,s,a) &= \max_{c,a' \geq 0} \Big\{ u(c) + \beta \sum_{s'} \pi_{ss'}\Big[(1-\epsilon_{wh}) \hat{U}^N(wh,s',a') + \epsilon_{wh} \hat{U}^N(wh,s',0)  \Big]\Big\}\\ 
        & \text{ s.t. }\\
        c + a' &= (1+r)a + d
\end{aligned}
\label{eqn:worker7}
\end{equation}
\Cref{eqn:worker7.2} expands the conditional future value of the white worker in this scenario. It is almost identical to \Cref{eqn:worker5.2} except for the fact that, in the event the unemployed worker doesn't find any job (with probability $(1-f(\theta_p))(1-f(\theta_{np}))$), the worker continues to remain ineligible for receiving unemployment benefits.

\begin{equation}
\begin{aligned}
        \hat{U}^N(wh,s',a') &=  (1-f(\theta_{np}))f(\theta_p) W_p(wh,s',a') +  f(\theta_{np})(1-f(\theta_p)) W_{np}(wh,s',a') \\  
        &+  f(\theta_p)f(\theta_{np})\max\{W_p(wh,s',a'),W_{np}(wh,s',a')\} \\         
        &+ (1-f(\theta_p))(1-f(\theta_{np})) U^N(wh,s',a') 
\end{aligned}
\label{eqn:worker7.2}
\end{equation}

\subsection{Firms}

A large number of $p$ and $np$ firms post vacancies after paying the vacancy posting costs. All firms are risk-neutral and discount their future profits using the equilibrium real interest rate. 

\subsubsection{Vacant \textit{np} firm}

Vacant $np$ firms pay a cost of $\kappa_{np}$ and search among all unemployed workers irrespective of their race. These firms match with an unemployed worker with probability $q(\theta_{np})$. The value of a vacant $np$ firm, $V_{np}$ is given by \Cref{eqn:firm1}.

\begin{equation}
\begin{aligned}
    V_{np} &= -\kappa_{np} + \frac{q(\theta_{np})}{1+r} \int_{a} \Bigg\{\sum_{s'} \pi_{ss'} \Big[ (1-\epsilon_{bl}) J_{np}(bl,s',g(bl,s,a)) + \epsilon_{bl} J_{np}(bl,s',0)\Big] \frac{\phi_u(bl,s,a)}{u} \\
    &+ \sum_{s'} \pi_{ss'} \Big[(1-\epsilon_{wh}) \tilde{V}_{np}(wh,s',g(wh,s,a)) + \epsilon_{wh} \tilde{V}_{np}(wh,s',0)\Big] \frac{\phi_u(wh,s,a)}{u}\Bigg\} da 
\end{aligned} 
\label{eqn:firm1}
\end{equation}
where
\begin{equation*}
\begin{aligned}
   \tilde{V}_{np}&(wh,s',g(wh,s,a)) = \mathbbm{1}_{\{W_{np}(wh,s',g(wh,s,a))\geq W_{p}(wh,s',g(wh,s,a))\}} \Big(J_{np}(wh,s',g(wh,s,a)) \Big) \\
   &+ \mathbbm{1}_{\{W_{np}(wh,s',g(wh,s,a))< W_{p}(wh,s',g(wh,s,a))\}} \Big( J_{np}(wh,s',g(wh,s,a))\Big)(1-f(\theta_p))
\end{aligned}    
\end{equation*}

with $g(R,s,a)$ denoting the worker's optimal choice of asset $a'$ for the next period. The firm's expected value of matching with a worker depends on the probability of matching with a specific worker of type $(R, s, a)$ and the expected value of production. An $np$ firm can match with either a black or a white worker from the current unemployment pool. $\phi_u(bl,s,a)$ is the mass of unemployed black workers with productivity $s$ and asset $a$, while $\phi_u(wh,s,a)$ is the corresponding mass of white workers. Thus, $\phi_u(bl,s,a)/u$ and $\phi_u(wh,s,a)/u$
reflect the probabilities that a vacant $np$ firm gets matched with an unemployed worker of type $(R, s, a)$. An unemployed black worker will accept the $np$ job offer once they get matched. On the other hand, since unemployed white workers can potentially receive a competing offer from a $p$ firm, the white worker will accept the $np$ job and begin producing only if the $np$ job provides the worker with a higher value, denoted by $\mathbbm{1}_{\{W_{np}(wh,s',g(wh,s,a))\geq W_{p}(wh,s',g(wh,s,a))\}}$, or if the worker did not receive a $p$ firm offer, described by $\mathbbm{1}_{\{W_{np}(wh,s',g(wh,s,a))< W_{p}(wh,s',g(wh,s,a))\}}(1-f(\theta_p))$. 

\subsubsection{Vacant \textit{p} firm}

Vacant $p$ firms pay a posting cost of $\kappa_p$ and restrict their search to unemployed white workers. The probability of matching with an unemployed white worker is $q(\theta_{p})$. \Cref{eqn:firm2} describes the value of maintaining a $p$ vacancy, $V_p$.

\begin{equation}
\begin{aligned}
    V_{p} &= -\kappa_{p} + \frac{q(\theta_{p})}{1+r} \int_{a} \Bigg\{ \sum_{s'} \pi_{ss'} \Big[(1-\epsilon_{wh}) \tilde{V}_{p}(wh,s',g(wh,s,a)) + \epsilon_{wh} \tilde{V}_{p}(wh,s',0)\Big] \frac{\phi_u(wh,s,a)}{u_{wh}}\Bigg\} da 
\end{aligned} 
\label{eqn:firm2}
\end{equation}
where
\begin{equation*}
\begin{aligned}
   \tilde{V}_{p}&(wh,s',g(wh,s,a)) = \mathbbm{1}_{\{W_{p}(wh,s',g(wh,s,a))> W_{np}(wh,s',g(wh,s,a))\}} \Big(J_{p}(wh,s',g(wh,s,a)) \Big) \\
   &+ \mathbbm{1}_{\{W_{p}(wh,s',g(wh,s,a))\leq W_{np}(wh,s',g(wh,s,a))\}} \Big( J_{p}(wh,s',g(wh,s,a))\Big)(1-f(\theta_{np}))
\end{aligned}    
\end{equation*}
Since the vacant $p$ firm searches only among white workers, the probability of matching with an unemployed white worker with productivity $s$ and asset $a$ is given by $\phi_u(wh,s,a)/u_{wh}$. Similar to the case of $np$ firms, a worker will accept the $p$ job only if the offer is more favorable than the other option or if they do not receive a competing offer from an $np$ firm.

We assume that there is free entry of firms, and therefore both $p$ and $np$ firms post vacancies until $V_p = 0$ and $V_{np} = 0$, respectively.

\subsubsection{Producing \textit{np} firm}

The problem of a producing $np$ firm is described by \Cref{eqn:firm3}. The firm chooses capital $k$ to maximize its lifetime value,  with the future value discounted using the interest rate and adjusted for the job destruction rate $\lambda_{np}$. Since the worker may experience an extreme wealth shock with probability $\epsilon_{R}$, it changes the wage bargaining position of the worker to the firm. The future value of the firm incorporates this probability. 

In the current period, \( j(R, s, a) \) represents the profit earned, calculated as the output net of production costs. The matched firm produces \( sf(k) \) units of output and incurs costs for renting capital and depreciation, denoted by \( (r + \delta)k \), as well as wage costs, \( \omega_{np}(R, s, a) \).

\begin{equation}
\begin{aligned}
        J_{np}(R,s,a) &= \max_k\Big\{j(R,s,a) + \frac{1-\lambda^{np}}{1+r} \sum_{s'} \pi_{ss'} \big[ (1-\epsilon_{R}) J_{np}(R,s',a') + \epsilon_{R} J_{np}(R,s',0) \big]\Big\}\\ 
        & \text{where}\\
        j(R,s,a) &= sf(k) - (r+\delta)k - \omega_{np}(R,s,a).
\end{aligned}
\label{eqn:firm3}
\end{equation}

\subsubsection{Producing \textit{p} firm}

Similar to a producing $np$ firm, \Cref{eqn:firm4} describes a producing $p$ firm. The firm rents capital and discounts future profit accounting for the job destruction rate $\lambda_p$ and the worker's extreme wealth shock $\epsilon_{wh}$.

\begin{equation}
\begin{aligned}
        J_{p}(wh,s,a) &= \max_k\Big\{j(wh,s,a) + \Big(\frac{1-\lambda^{p}}{1+r}\Big) \sum_{s'} \pi_{ss'} \big[ (1-\epsilon_{wh})J_{p}(wh,s',a') + \epsilon_{wh}J_{p}(wh,s',0) \big] \Big\}\\ 
        & \text{where}\\
        j(wh,s,a) &= sf(k) - (r+\delta)k - \omega_{p}(wh,s,a).
\end{aligned}
\label{eqn:firm4}
\end{equation}

\subsection{Wage bargaining}

In each period, matched worker-firm pairs bargain for wages over the match surplus. A worker upon matching earns a surplus of $(W_{i}(R,s,a)-U^I_{i}(R,s,a))$ while the firm surplus is given by $J_{i}(Ra,s,a)$, where $i$ denotes $np$ or $p$.  The bargaining power of the worker $\xi_{R}$ depends on the individual's race, and the firm's bargaining power is $1-\xi_{R}$. The resulting wage for workers employed at $p$ and $np$ firms are given by \Cref{eqn:wagenp} and \Cref{eqn:wagep}, respectively. 

\begin{equation}
\begin{aligned}
\omega_{np}(R,s,a) &= \argmax_{\omega_{np}} \Big(W_{np}(R,s,a)-U^I_{np}(R,s,a)\Big)^{\xi_{R}} J_{np}(Ra,s,a)^{1-\xi_{R}}
\end{aligned}
\label{eqn:wagenp}
\end{equation}

\begin{equation}
\begin{aligned}
\omega_{p}(wh,s,a) &= \argmax_{\omega_{p}} (W_{p}(wh,s,a)-U^I_{p}(wh,s,a))^{\xi_{wh}} J_{p}(wh,s,a)^{1-\xi_{wh}}
\end{aligned}
\label{eqn:wagep}
\end{equation}
We define the stationary equilibrium in \ref{appA}. 

\section{Calibration}\label{sec: cali}

Each period in the model represents a quarter. We calibrate the benchmark model to match the relevant US economy moments. We have two sets of parameters. One group of parameters is chosen externally based on the literature and empirical evidence without using model-generated data. The other set of parameters is calibrated internally by simulating our model to match a set of relevant data moments. \Cref{tab: calibration} shows the values of the internal calibrated parameters, their targets, along with the externally chosen parameters and their sources. We set the percentage of black workers in the model at 14.9\%, based on the Panel Study of Income Dynamics (PSID) data from 1996 to 2014.\footnote{Our sample follows \cite{blundell2018children}. The share calculation follows \citet{nakajima2021monetary}.}

\begin{table}[htbp]
  \centering
  \caption{Calibration and targeted statistics}
 \scalebox{0.8}{

\begin{tabular}{llllcc}\toprule
Parameter & Value & Description & Target statistics & data & model \\\midrule 
\\
\multicolumn{6}{l}{\textit{Chosen internally}} \\ 
\\
$\beta $       & 0.9943 &  discount factor  & K/Y             & 10.26   & 10.26 \\
$\iota$      & 1.3452 & matching elasticity                 & job finding rate - black       & 0.4946 & 0.4946 \\
$\kappa_{p}$  & 4.6314 & p sector vacancy posting cost       & job finding rate - white       & 0.6599 & 0.6599 \\
$\kappa_{\text{np}}$ & 2.5534 & np sector vacancy posting cost & market tightness & 1.0      & 1.0 \\               
$\lambda_{np}$& 0.0644 & np sector job destruction rate     & job separation rate - black    & 0.0644 & 0.0644 \\
$\lambda_{p}$ & 0.0268 & p sector job destruction rate      & job separation rate - white    & 0.0380 & 0.0380 \\
$\xi_{bl}$    & 0.1371 & bargaining power - black            & mean wage ratio              & 0.75 & 0.75 \\
$\xi_{wh}$    & 0.1988 & bargaining power - white            & firm profit share              & 0.033  & 0.033 \\
$\epsilon_{bl}$    & 0.0179 & extreme wealth shock - black            & zero wealth share - black              & 0.18 & 0.18 \\
$\epsilon_{wh}$    & 0.0088 & extreme wealth shock - white            & zero wealth share - white              & 0.07  & 0.07 \\
$\chi$   & 0.8403 & maximum UI coverage & fraction of median wage & 0.48 & 0.48 \\
$Pe$  & 0.5385 & prob. of UI eligibility & average weeks of eligibility & 26 & 26 \\ \hline
\\
\multicolumn{6}{l}{\textit{Chosen externally}} \\ 
\\
$\alpha_k$ & 0.2890 & capital share of output & \citet{nakajima2012business} &       &  \\
$\delta$ & 0.0150 & quarterly depreciation rate & \citet{nakajima2012business}  &    &   \\
$\rho_s$   & 0.9411 & persistence of idiosyncratic shock  &  PSID &       &  \\
$\sigma_s$ & 0.1680 & standard deviation of idiosyncratic shock & PSID &       &  \\
$h$     & 0.4000   & UI replacement rate & \citet{mitman2015optimal} &       &  
\\\bottomrule
\end{tabular}%
}

 \begin{minipage}{\linewidth}{\centering \scriptsize \singlespacing Notes: This table presents the model parameters, their values, and descriptions. The top panel displays the parameters selected internally by minimizing the distance between the moments generated by the model and the moments derived from the data. The last two columns of the top panel compare the targeted moments in the data with those from the model simulations. The bottom panel provides the parameters chosen externally, their values, their descriptions and sources.} 
     \end{minipage}    
  \label{tab: calibration}%
\end{table}%

\subsection{Preferences}

We set the period utility function $u(c)$ to be $\log (c)$. The discount factor, $\beta$, is calibrated to match the quarterly capital-output ratio of 10.26, a value used by a number of studies such as \cite{den2010computational} and \cite{carroll2017distribution}. The resulting value of $\beta$ is 0.9943, and the corresponding quarterly real interest rate is 1.3\%. 

\subsection{Production}
The matched worker-firm pairs produce according to a Cobb-Douglas production function, $f(k) = k^{\alpha_k}$. We choose $\alpha_k$ to be 0.289 and set the quarterly capital depreciation rate $\delta$ equal to 0.015, following \cite{nakajima2012business}. 

\subsection{Productivity and wealth shocks}
We use hourly real wages from the PSID to estimate the persistence, $\rho_s$, and the standard deviation, $\sigma_s$ of the productivity process. Our estimation strategy follows \cite{setty2021provision}. We run a Mincer regression of log wages on demographic controls and year and state fixed effects. The demographic controls include education, labor market experience, race, and marital status. We restrict the sample to men aged 25 and older. We then use the residuals obtained to estimate an AR(1) regression using the panel dimension of the PSID data. The estimated regression operates on a biennial frequency since the PSID data is released every two years. We assign the quarterly adjusted coefficient of the AR(1) regression as the persistence of the idiosyncratic productivity process. The standard deviation of the productivity process corresponds to the standard deviation of the residuals from the AR(1) regression after adjusting for the model frequency. We estimate the quarterly persistence, $\rho_s$, to be 0.9411 and the corresponding standard deviation, $\sigma_s$, to be 0.1680.\footnote{We follow \citet{setty2021provision} to have only men in the sample. We also estimate the AR(1) process to include both men and women from PSID. The resulting $\rho_s$ is unchanged (0.9411), and $\sigma_s$ is slightly lower (0.1671) than the men-only sample.}

Following \cite{mukoyama2010welfare}, our model also features extreme wealth shocks to capture the mass of people with zero wealth. The race-specific probability of losing one's wealth, $\epsilon_R$, is calibrated to be 0.0179 for black workers and 0.0088 for white workers. They capture the empirical moments that around 18\% of black workers and 7\% of white workers have zero wealth \citep{nakajima2021monetary}.

\subsection{Unemployment insurance}

The unemployment insurance system in our model is characterized by the replacement rate, $h$, the maximum insurance payout, $\chi$, and the probability of maintaining eligibility status, $P_e$. We choose $h$ to be 0.4 following \citet{shimer2005cyclical} and \citet{mitman2015optimal}. In accordance with \citet{setty2021provision}, we calibrate $\chi$ to 0.8403, which amounts to 48\% of the median wage in the model. The eligibility probability $P_e$ is set to 0.5385 to generate an average duration of unemployment benefits of 26 weeks, as in \citet{mitman2015optimal}.

\subsection{Labor search and wage bargain}
The parameters governing the labor search are calibrated to match the labor market turnover statistics obtained from \citet{cajner2017racial}. The elasticity of the matching function, $\iota$, targets the job-finding rate of black workers. Our calibrated value of 1.3452 is close to 1.25, the value used by \cite{den2000job} and \cite{petrosky2018endogenous}. We choose firm-type specific job destruction rates $\lambda_{np}$ and $\lambda_{p}$ to match the job separation rates of black and white workers, respectively. The ensuing values for $\lambda_{np}$ and $\lambda_{p}$ are 0.0644 and 0.0268, respectively.\footnote{All the employed black workers only work with $np$ firms, and hence $\lambda_{np}$ is exactly equal to the job separation rate of black workers. White workers are employed in both $p$ and $np$ firms. Hence, $\lambda_p$ (0.0268) is lower than the aggregate separation rate of white workers (0.0380).} The vacancy posting cost of prejudiced firms, $\kappa_p$, is chosen to match the job-finding rate of white workers. The posting cost of non-prejudiced firms, $\kappa_{np}$, is chosen to match the aggregate labor market tightness $\theta$, following \citet{wolcott2021employment}.\footnote{Aggregate labor market tightness is defined as $\theta = \frac{v_{np}+v_p}{u_{bl}+u_{wh}}$.} We find that prejudiced firms pay significantly more than non-prejudiced firms to post their vacancies, with $\kappa_p$ calibrated at 4.6314, compared to $\kappa_{np}$ at 2.5534. The bargaining power of black workers, $\xi_{bl}$, is calibrated to target the average black-white racial wage ratio of 0.75 \citep{derenoncourt2021minimum}. The bargaining power of the white worker, $\xi_{wh}$, is chosen to generate the average firm profit share of 3.3\% as in \cite{nakajima2012business}. Consistent with our expectations, black workers have a lower bargaining power than white workers, with $\xi_{bl}$ calibrated to be 0.1371 and $\xi_{wh}$ to be 0.1988. 

\subsection{Assessing the model as a quantitative theory of racial disparity}

Our calibration successfully captures the racial gaps across three important dimensions: income, wealth, and employment outcomes. We now discuss the model validity in more detail before proceeding with the quantitative exercises that examine the macroeconomic impact of racial discrimination.

\begin{figure}
    \centering
    \begin{subfigure}{0.5\linewidth}
        \centering
        \includegraphics[width=\linewidth]{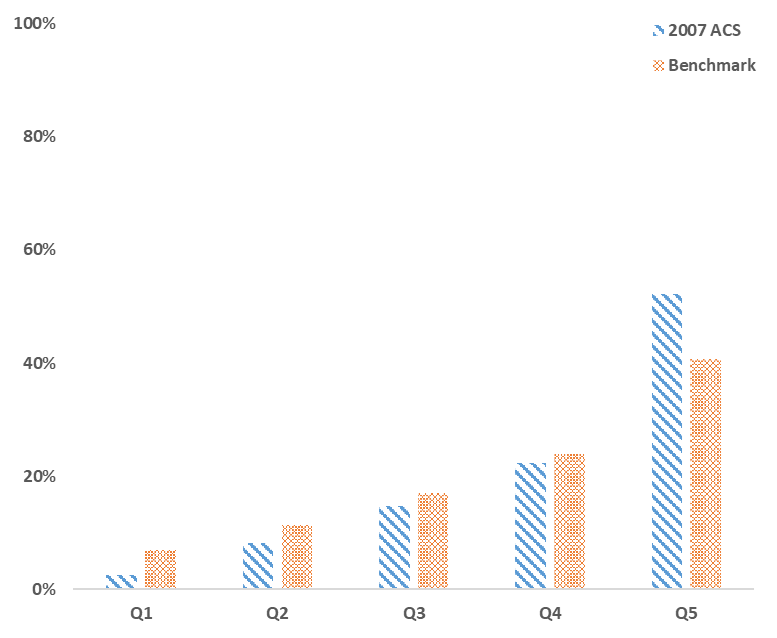}
        \caption{Labor income share}
        \label{fig:earnings_share}
    \end{subfigure}%
    \begin{subfigure}{0.5\linewidth}
        \centering
        \includegraphics[width=\linewidth]{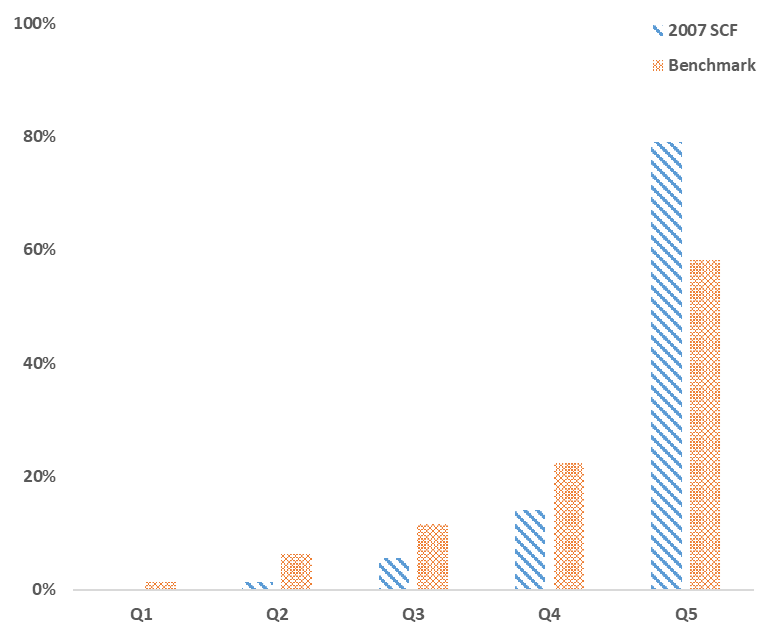}
        \caption{Wealth share}
        \label{fig:wealth_share}
    \end{subfigure}
     \caption{Labor income and wealth distributions}

\vspace{3mm}
    
    \begin{minipage}{0.99\textwidth} 
{\footnotesize \textit{Note:} This figure compares the steady-state distributions of wealth and labor income generated by the model with their empirical counterparts. The empirical labor income and wealth distributions are estimated using data from the 2007 American Community Survey (ACS) and the 2007 Consumer Finance Survey (SCF), respectively. The horizontal axes indicate quintiles, from the lowest (Q1) to the highest (Q5), while the vertical axes report the share of total income or wealth accounted for by each quintile. Blue bars represent the data, and red dotted bars show the model-simulated moments.}
\end{minipage}
    \label{fig:distribution}
\end{figure}

First, the model successfully captures the overall household distribution of wealth and labor earnings as shown in \Cref{fig:distribution}, even though we do not explicitly target these distributions. As with most incomplete market models, we also face difficulty in generating the extreme concentration of wealth and labor income in the top quintile of the distribution. On the other hand, we are much closer to the empirical distributions in the lower quintiles since we target the share of zero-wealth workers in our calibration.

\begin{table}
\centering
\small
\caption{Steady state racial inequality}
\begin{tabular}{lcc}\toprule
Moments & Data & Model \\\midrule
Unemp rate (Black) & 0.12 & 0.12 \\
Unemp rate (White) & 0.05 & 0.05 \\
Mean wealth ratio & 0.23 & 0.29 \\
Median wealth ratio & 0.17 & 0.33 \\
\bottomrule
\end{tabular}%
\vspace{3mm}
    
    \begin{minipage}{0.99\textwidth} 
{\footnotesize \textit{Note:} This table compares the empirical unemployment and wealth moments with the corresponding steady-state moments generated by the model.}
\end{minipage}
\label{tab:ss_benchmark}
\end{table}

Second, the model reproduces the disparities in unemployment, labor income, and wealth between black and white workers. Through calibration, the model replicates the empirical racial labor income gap and the empirical racial difference in the share of people at zero wealth. In addition to the targeted moments in calibration, \Cref{tab:ss_benchmark} shows that our model successfully generates the untargeted employment and wealth moments close to the data. Specifically, our model produces a 5\% unemployment rate among white workers compared to 12\% among black workers. Apart from the lower separation rate, white workers have a lower unemployment rate due to their access to prejudiced firms. On the wealth dimension, without targeting, the model generates a mean black-white wealth ratio of 0.29, close to 0.23 in the data \citep{kaplan2014wealthy}. However, the model underestimates the median wealth gap between black and white workers. The median black worker holds 33\% of the wealth of the median white worker in our model compared to the empirical value of 17\% \citep{kaplan2014wealthy}. 

In summary, the model reflects the racial disparities present in labor income, wealth, and unemployment outcomes. Next, we investigate how racial discrimination in hiring contributes to these disparities.

\section{Steady state results}\label{sec: steady}

In this section, we first establish how a search framework sustains hiring discrimination in equilibrium.  Then, we examine the steady-state impact of racial discrimination by comparing the benchmark economy with an alternative economy without hiring discrimination. Lastly, we explore the heterogeneous welfare implications associated with eliminating hiring discrimination.

\subsection{Sustaining racial discrimination}

Search-and-matching frictions generate match surplus that firms can partially appropriate, allowing discriminatory and non-discriminatory vacancies to coexist with positive profits in equilibrium. This contrasts with the frictionless benchmark, in which taste-based discrimination is competed away when groups are equally productive, as in \citet{becker19571971}. More broadly, general equilibrium models show that search frictions and adjustment costs can sustain persistent firm behavior that deviates from the frictionless optimum and can give rise to inefficiencies and welfare losses even in the absence of underlying productivity differences \citep{hopenhayn1993job, cooper2007search, yang2018welfare}.

\begin{table}[htbp]
\centering
\small
\caption{Firms in the steady state}
\begin{tabular}{lc}
\toprule
Moments & Benchmark \\
\midrule
$p$ firm profit & 0.04 \\
\textit{np} firm profit & 0.02 \\[0.5em]
\textit{p} firm vacancy & 0.02 \\ 
\textit{np} firm vacancy & 0.05 \\[0.5em] 
\textit{p} firm employment & 0.57 \\ 
\textit{np} firm employment & 0.37 \\
\bottomrule
\end{tabular}
\vspace{3mm}

    \begin{minipage}{0.99\textwidth} 
{\footnotesize \textit{Note:} This table presents steady state profit, vacancy postings, and employment levels for $p$ and $np$ firms in the benchmark model.}
\end{minipage}
\label{tab:firms}
\end{table}

\Cref{tab:firms} shows the performance of firms in the steady state. Prejudiced firms earn a total profit of 0.04, while the corresponding profit of non-prejudiced firms is 0.02. This difference arises because $p$ firms experience a lower job destruction shock (0.0268) compared to $np$ firms (0.0644). Even though $p$ firms incur higher vacancy posting costs (4.6314) to hire only white workers, they retain their workers for a longer period, reducing the number of vacancy postings. In contrast, $np$ firms face lower vacancy posting costs (2.5534), but their higher job destruction rate leads to a more frequent need for vacancy postings (0.05). In the steady state, $p$ firms sustain an employment level (0.57) higher than $np$ firms (0.37).

\subsection{Impact of hiring discrimination on racial inequality}

In this subsection, we analyze the distributional and aggregate effects of discriminatory hiring by comparing the benchmark economy to a counterfactual economy without prejudiced firms. The counterfactual economy contains only non-prejudiced firms, where both black and white workers retain their race-specific job separation rates from the benchmark calibration. By holding these separation rates constant, we isolate how discrimination in hiring - rather than differences in job separation - shapes labor market outcomes.\footnote{The benchmark economy calibrates prejudiced and non-prejudiced firms' job destruction shocks ($\lambda_{p}$ and $\lambda_{np}$) to match observed racial disparities in job separation rates. In the single-firm-type counterfactual, we preserve benchmark separation rates by assigning white workers a $\lambda_{wh}$ of 0.0380 and black workers a $\lambda_{bl}$ of 0.0644.} \Cref{tab:penalize} presents the comparison.

\begin{table}
\centering
\small
\caption{Impact of hiring discrimination}

\begin{tabular}{lcc}
\toprule
Moments & Benchmark & No $p$ firms \\
\midrule
\multicolumn{3}{c}{\emph{Households}} \\
job separation rate - black & 0.06 & 0.06 \\
job separation rate - white & 0.04 & 0.04 \\[0.5em]
job finding rate - black    & 0.49 & 0.64 \\
job finding rate - white    & 0.66 & 0.64 \\[0.5em]
unemp rate - black          & 0.12 & 0.09 \\
unemp rate - white          & 0.05 & 0.06 \\[0.5em]
mean wage - black           & 1.59 & 1.85 \\
mean wage - white           & 2.12 & 2.11 \\
mean wage ratio             & 0.75 & 0.87 \\[0.5em]
mean wealth ratio           & 0.29 & 0.34 \\
median wealth ratio         & 0.33 & 0.44 \\
\midrule
\multicolumn{3}{c}{\emph{Firms}} \\
$p$ firm profit           & 0.04 & -    \\
\textit{np} firm profit     & 0.02 & 0.06 \\[0.5em]
$p$ firm vacancy          & 0.02 & -    \\
\textit{np} firm vacancy    & 0.05 & 0.07 \\[0.5em]
$p$ firm employment       & 0.57 & -    \\
\textit{np} firm employment & 0.37 & 0.94 \\
\midrule
\multicolumn{3}{c}{\emph{Labor Market}} \\
$p$ market tightness      & 0.39  & -    \\
$np$ market tightness     & 0.71 & 1.17 \\
\midrule
\multicolumn{3}{c}{\emph{Aggregate Outcomes}} \\
Y                           & 3.07 & 3.08 \\
K/Y                         & 10.26 & 10.27 \\
average wage                & 2.04 & 2.07 \\
unemp rate                  & 0.06 & 0.06 \\
\bottomrule
\end{tabular}

\vspace{3mm}
    
    \begin{minipage}{0.99\textwidth} 
{\footnotesize \textit{Note:} This table compares the benchmark steady state with a counterfactual steady state without $p$ firms. In the counterfactual model, all firms provide equal hiring opportunities to black and white workers. All other parameters stay the same as benchmark calibration.}
\end{minipage}
\label{tab:penalize}
\end{table}

The first panel shows how racial disparities in household outcomes change when prejudiced firms are removed. While the racial differences in job separation rates remain unchanged at benchmark levels, eliminating discriminatory hiring practices leads to equalized job-finding rates for black and white workers. White workers experience a slight decrease in their job-finding rate, dropping from 0.66 to 0.64, due to losing access to $p$ firms. Meanwhile, black workers' job-finding rate rises significantly, from 0.49 to 0.64, in the counterfactual equilibrium. The removal of $p$ firms leads to endogenous entry of $np$ firms, which increases the number of vacancies and thus raises the job-finding rate for black workers. Overall, this removal in prejudiced firms narrows the racial unemployment gap from 7 percentage points to 3 percentage points, showing that discriminatory hiring explains approximately 57\% of the unemployment gap between black and white workers.

Turning to the impact on wages, removing hiring discrimination boosts outcomes for black workers by improving their outside options. This enables black workers to bargain for higher wages upon a successful match. However, white workers no longer have access to jobs from the exclusive $p$ firms. This lowers their outside option during bargaining, slightly reducing their wage rates. As a result, the average wage ratio between black and white workers increases from 0.75 to 0.87. We can conclude that hiring discrimination accounts for roughly 48\% of the wage gap.

The change in hiring practices also narrows the steady-state wealth gap: the mean black-to-white wealth ratio increases from 0.29 to 0.33, and the median ratio rises from 0.34 to 0.44. These changes imply that discriminatory hiring accounts for roughly 7\% of the mean wealth gap and 16\% of the median wealth gap. The remaining wage and wealth differences reflect disparities in bargaining power and the incidence of extreme idiosyncratic wealth shocks between black and white workers.\footnote{Although not the primary focus of this paper, additional results that remove racial differences in bargaining power and extreme wealth shocks are presented in \ref{appB}.} This result is consistent with work showing that persistent differences in labor market risk can translate into long-run wealth inequality through precautionary saving and nonlinear accumulation in incomplete-markets general equilibrium models \citep{ganong2020wealth, aliprantis2023dynamics, yang2025better}.

The second panel examines firm dynamics when discriminatory firms are removed.\footnote{We refer to the one-sector economy without discriminatory hiring as $np$ firms, although their job destruction rate $\lambda$ differs from that of the benchmark $np$ firms.} Without prejudiced firms, $np$ firms see their profits triple, which encourages additional $np$ firms to enter the market and post vacancies. Consequently, $np$ firms fully replace $p$ firms in the counterfactual equilibrium. As shown in \Cref{tab:penalize}, the total number of vacancies and overall employment remain unchanged compared to the benchmark. Still, eliminating prejudiced firms helps narrow the racial gap in labor market outcomes, as described earlier.

The third panel of \Cref{tab:penalize} shows how labor market conditions adjust when $p$ firms exit. As these firms leave, additional $np$ firms enter endogenously, raising the labor market tightness in the $np$ market from 0.71 in the benchmark to 1.17 in the counterfactual equilibrium. Consequently, the overall market tightness exceeds the benchmark aggregate level of 1.

The last panel of \Cref{tab:penalize} presents the aggregate outcomes. Despite the increased entry of $np$ firms, aggregate output rises only marginally from 3.07 to 3.08. The capital-output ratio increases from 10.26 to 10.27, and the aggregate wage grows from 2.04 to 2.07, while the aggregate unemployment rate remains steady at 6\%.

Our results suggest that prejudiced firms account for a substantial share of employment, income, and wealth inequality. Eliminating $p$ firms improves employment and wage outcomes for black workers, while white workers experience a slight decline in their job-finding rate and an increase in unemployment. This reduction in job prospects weakens white workers' bargaining power and adversely affects their income and wealth. In the next section, we quantify these differential effects by calculating the welfare changes experienced by black and white workers.

\subsection{Welfare analysis}\label{sec: welfare}

We follow \citet{krusell2010labour} in measuring the welfare effects of discriminatory hiring on black and white workers. To do so, we compute the change in average consumption equivalence after removing $p$ firms.\footnote{See \ref{appC} for details on the welfare calculation.} \Cref{tab: welfare} presents the overall welfare change, as well as the breakdown by productivity type and wealth quintile for both groups.

Eliminating hiring discrimination raises aggregate welfare by 1.5\%. Black workers experience a 12.9\% increase in their average welfare, mainly due to improved job opportunities and higher wage rates. White workers, on the other hand, experience a modest 0.48\% decline, driven by a slight drop in wages and a higher unemployment rate after losing access to the exclusive $p$ firms. Although white workers make up about 85\% of the population, the substantial gains for black workers more than offset the small loss among white workers, resulting in a net rise in overall welfare. These results are consistent with a broader class of general equilibrium models with labor market frictions, in which relatively modest changes in aggregate quantities can translate into sizable welfare effects through changes in risk exposure and match stability \citep{hopenhayn1993job, yang2018welfare}.

\begin{table}[htbp]
  \centering
    \caption{Heterogeneous welfare change}
\begin{tabular}{lcc}
\toprule
Average welfare gain (\%) & \multicolumn{2}{c}{No $p$ firms} \\[0.5em]
\hline
\\
Aggregate & \multicolumn{2}{c}{1.51} \\[0.5em]
\\
 & Black & White \\ \hline
 \\
Overall  & 12.89 & -0.48 \\[0.5em]
 \multicolumn{3}{c}{\emph{by productivity}} \\[0.5em]
Low   & 12.11 & -0.47 \\
Mid   & 12.94 & -0.48 \\
High  & 13.57 & -0.47 \\[0.5em]
 \multicolumn{3}{c}{\emph{by wealth}} \\[0.5em]
Low 20\% & 12.91 & -0.45 \\
40-60\%  & 12.92 & -0.49 \\
Top 20\% & 12.65 & -0.48 \\
\bottomrule
\end{tabular}

\vspace{3mm}

    \begin{minipage}{0.99\textwidth} 
{\footnotesize \textit{Note:} This table compares the average consumption equivalence change from the benchmark steady state to an equilibrium with no $p$ firms. Low, mid, and high productivity corresponds to the lowest, middle, and highest value of idiosyncratic productivity $s$. The wealth quintiles are based on benchmark steady-state wealth distribution.}
\end{minipage}
\label{tab: welfare}
\end{table}%

The middle panel in \Cref{tab: welfare} displays the heterogeneous welfare changes for black and white workers by productivity. For white workers, the welfare decline from eliminating prejudiced firms is similar across productivity levels, with a slightly larger loss for those in the middle productivity group. In contrast, among black workers, the highest productivity group experiences the greatest gain (13.57\%), followed by the middle and low productivity groups. The bottom panel in \Cref{tab: welfare} breaks down the welfare changes by wealth quintile. Removing prejudiced firms benefits black workers most in the middle quintile, where welfare rises by 12.92\%, while the largest welfare decline for white workers also occurs in the middle quintile.

\section{Business cycle dynamics}\label{sec: biz}

In this section, we first document the racial disparities over the business cycle. We then introduce aggregate uncertainty into our benchmark model to assess how hiring discrimination shapes racially disparate dynamics during economic fluctuations.

\subsection{Racial disparities over the business cycles}

Table \ref{tab:bizdata} displays the labor market outcome disparities between black and white workers over the business cycle. The top panel presents the dynamics of unemployment rates for both groups, while the bottom panel shows the evolution of average wages after adjusting for composition bias. In this table, volatility is defined as the standard deviation of a variable relative to that of real GDP, and cyclicality is measured as the correlation between the variable and real GDP.

\textbf{Unemployment rate:} We use quarterly data from the Current Population Survey (CPS) from 1996 to 2014 to calculate the business cycle dynamics of unemployment rates. The statistics are computed using log deviations from an HP-filtered trend with a smoothing parameter of 1600. We find that both black and white unemployment rates are countercyclical, although the black unemployment rate is slightly less cyclical and less volatile than the white rate. These findings align with \citet{nakajima2021monetary}. In addition, the black-white unemployment rate gap is itself countercyclical: it widens during recessions and narrows during expansions, indicating that black workers see a larger unemployment increase in recessions but a faster decrease in expansions. This result is consistent with the countercyclical gap documented by \citet{cajner2017racial}.

\begin{table}[htbp]
\centering
\caption{Business cycle statistics, US data, 1996-2014}
    \begin{tabular}{lcc}
    \toprule
         &  Volatility  &  Cyclicality  \\
    \midrule
    \multicolumn{3}{c}{\textit{Unemployment rate}} \\
    Black & 9.51 & -0.85 \\
    White & 11.02 & -0.92 \\
    Black-White gap & 9.46 & -0.63 \\
    \midrule
    \multicolumn{3}{c}{\textit{Average wage}} \\
    Black & 2.54 & 2.60 \\
    White & 2.08 & 1.27 \\
    \bottomrule
    \end{tabular}
\vspace{3mm}

    \begin{minipage}{0.99\textwidth} 
{\footnotesize \textit{Note:} This table provides business cycle statistics for the US. The unemployment rate is obtained from the Current Population Survey (CPS) for 1996 to 2014. Average wage statistics are constructed from the Panel Study of Income Dynamics (PSID) for the same period. Unemployment rate statistics are calculated using HP filtered series with a smoothing parameter of 1600, while wage statistics are obtained using wage growth rates after controlling for demographics.}
\end{minipage}
    \label{tab:bizdata}
\end{table}

 \textbf{Wages:} We use individual-level data from the PSID (1996--2014) to estimate the cyclical properties of black and white wages. Previous research \citep{stockman1983aggregation,bils1985real,solon1994measuring} shows that using aggregate data can be problematic due to changes in worker composition over the business cycle. Specifically, aggregate data tends to weight lower-skilled (and lower-wage) workers more heavily during expansions than recessions, causing countercyclical bias. To address this issue, we rely on the longitudinal structure of the PSID data to keep the composition of workers fixed over time. Following \citet{bils1985real}, \citet{solon1994measuring}, and \citet{devereux2001cyclicality}, we estimate wage cyclicality by regressing wage growth on real GDP growth, controlling for observable characteristics. We measure wage volatility as the standard deviation of wage growth after controlling for demographics.\footnote{\ref{appD} provides details on measuring the cyclicality and volatility of real wages.} Our findings indicate that both black and white wages are procyclical, with black workers' wages being twice as cyclical as those of white workers. Black workers' wages also exhibit greater volatility than white workers' wages.

\subsection{Augmented model with aggregate shocks}

We introduce shocks to aggregate total factor productivity (TFP) $z$ in our benchmark steady-state model described in \Cref{sec: model}. The TFP shocks follow an AR(1) process, $\log(z') = \rho_z \log(z) + \epsilon_z$, with $\epsilon_z \stackrel{iid}{\sim} N(0,\sigma_{z}^2)$. The output of the matched worker-firm pairs depends on the realizations of both aggregate and idiosyncratic productivity shocks and is given by $zsf(k)$.     

With the introduction of aggregate shocks, the state space expands to include aggregate states $(z,\mu)$, where $\mu$ is the distribution of workers over their employment status ($e$), race ($R$), idiosyncratic productivity ($s$), and asset ($a$). The aggregate distribution for the next period $\mu'$ is determined by $(z,\mu)$, and the endogenous law of motion given by $\mu' = \Gamma(z,\mu)$ is determined in the equilibrium.

We follow \citet{cooley1995economic} and \citet{boppart2018exploiting} in choosing the persistence parameter $\rho_z$ to be 0.95, while the standard deviation $\sigma_z$ is set to 0.015. All other parameters remain the same as our benchmark calibration given in \Cref{tab: calibration}. We obtain the stochastic equilibrium of our model using the sequence space method proposed by \citet{boppart2018exploiting}. We first solve non-linearly for the perfect foresight transitions to a single small MIT shock, i.e., an unexpected shock to the aggregate TFP. We then use the solved impulse responses as the numerical derivatives with respect to the initial TFP shock. Using these derivatives, we simulate the stochastic equilibrium by generating the TFP realizations and calculating the corresponding model moments over the business cycles as a linear combination of the impulse response and the TFP realizations.\footnote{This method hinges on the assumption that the business cycle dynamics can be well approximated as a linear system. In \ref{appE}, we demonstrate the validity of this assumption by establishing the symmetry of the impulse responses to 1\% positive and negative TFP shocks.}

\begin{table}[htbp]
\centering
\caption{Business cycle model vs data} 
\begin{tabular}{lcccc}
\toprule
& \multicolumn{2}{c}{Volatility} & \multicolumn{2}{c}{Cyclicality} \\
 & Data & Model & Data & Model \\
[0.5em] \hline \\ 
& \multicolumn{4}{c}{\textit{Unemployment rate}} \\
[0.5em]
Black & 9.51 & 0.26 & -0.85 & -0.69 \\
White & 11.02 & 0.22 & -0.92 & -0.71 \\
Black-White gap & 9.46 & 0.30 & -0.63 & -0.68 \\
[0.5em]
& \multicolumn{4}{c}{\textit{Average wage}} \\
[0.5em] 
Black & 2.54 & 1.08 & 2.60 & 1.08 \\
White & 2.08 & 1.00 & 1.27 & 1.00 \\
\bottomrule
\end{tabular}

\vspace{3mm}
    
    \begin{minipage}{0.99\textwidth} 
{\footnotesize \textit{Note:} This table compares the data moments with the moments generated by the benchmark model on unemployment and wages.}
\end{minipage}
\label{tab: cyclegaps_model_data} 
\end{table}

\Cref{tab: cyclegaps_model_data} compares business cycle statistics for unemployment and wages between the data and the model. Our model generates countercyclical unemployment rates, with white unemployment being slightly more cyclical than black unemployment, and it accurately reproduces the countercyclical gap between the two groups. In terms of wages, the model produces procyclical patterns, with black wages showing higher cyclicality and volatility than white wages. However, as is common with standard search models, our model falls short of generating realistic volatility levels for both black and white unemployment rates - a limitation discussed in \citet{shimer2005cyclical}.

\subsection{Impact of hiring discrimination over business cycle}

We compare the benchmark dynamics with a counterfactual model that excludes prejudiced firms to examine how discriminatory hiring impacts the economy under aggregate fluctuations. In this one-sector counterfactual, only non-prejudiced firms operate, and workers face race-specific job separation rates as with \Cref{tab:penalize}. 

\subsubsection{Impulse responses}

\begin{figure}[htbp]
\centering
\includegraphics[width=\textwidth]
{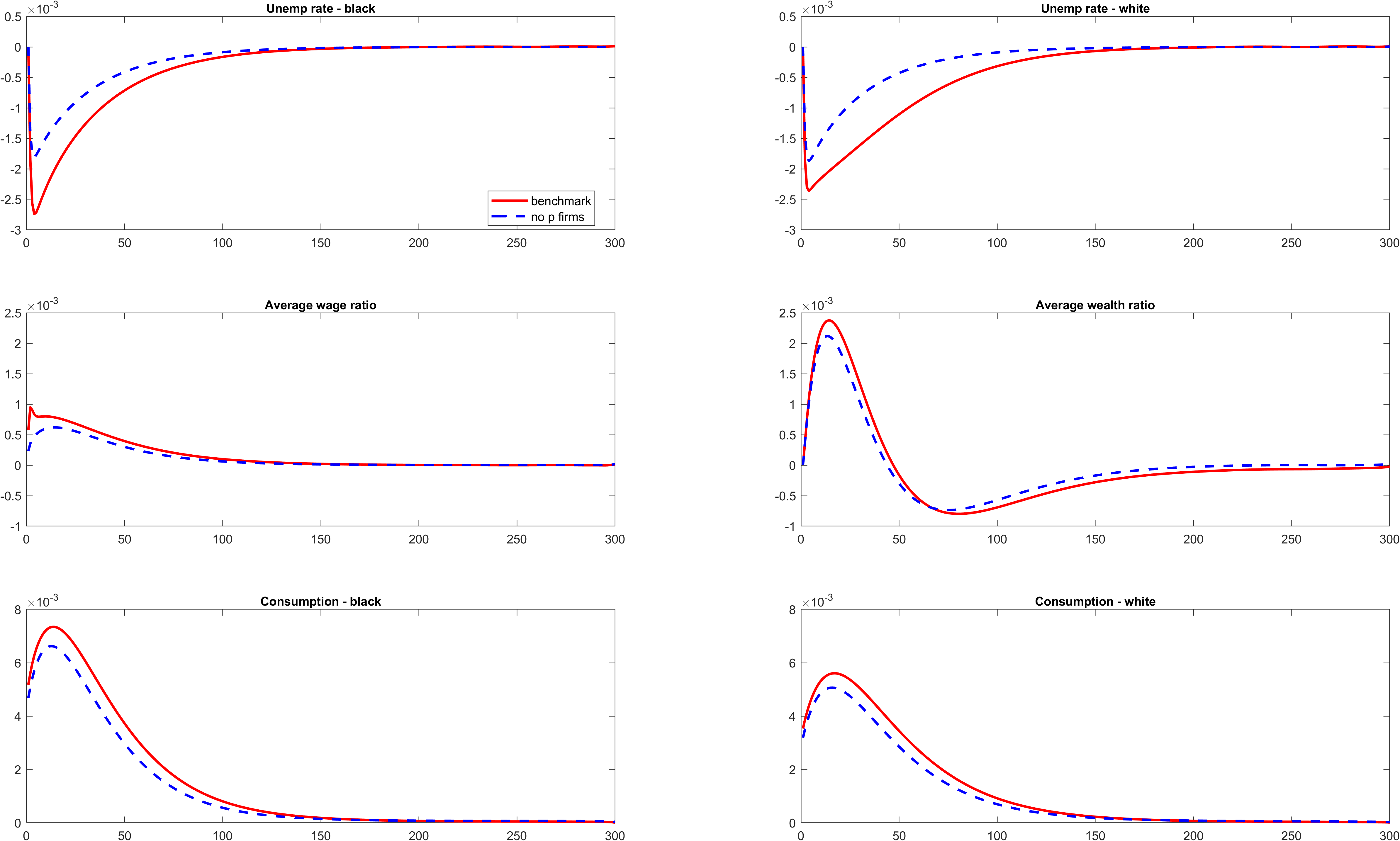}
\caption{Impulse responses}
\vspace{3mm}

    \begin{minipage}{0.99\textwidth} 
{\footnotesize \textit{Note:} This figure plots the impulse response functions of unemployment rates, black-white average wage and wealth ratios, and aggregate consumption of black and white workers to a one percent increase in aggregate TFP at date zero. The red solid line denotes the responses from the benchmark model, and the blue dashed line denotes the responses from the counterfactual model without $p$ firms.}
\end{minipage}
\label{fig:irf}
\end{figure}

\Cref{fig:irf} shows how labor market outcomes, wealth, and consumption for both black and white workers respond to an unexpected 1\% expansionary TFP shock. We compare the benchmark model's reaction with that of a counterfactual model that excludes $p$ firms. In both models, unemployment rates move countercyclically while wages, wealth, and consumption follow procyclical patterns. However, the responses in the counterfactual economy are generally less pronounced.

In the model without $p$ firms, the unemployment rate for black workers falls less after an expansionary shock than in the benchmark, indicating that black unemployment is less volatile when discriminatory hiring is removed. In the benchmark economy, $np$ firms compete with $p$ firms for hires, which leads them to post more vacancies and create additional opportunities for workers. Although white workers follow a similar pattern, the difference between the benchmark and counterfactual responses is much smaller for them. Notably, the counterfactual model shows a much faster recovery of unemployment rates than the benchmark model.\footnote{This is a common feature in dual labor market models, where a single-market setup typically reverts to its steady state faster than a dual-market framework \citep[e.g.,][]{horvath2022unemployment}.}

The average black-white wage and wealth ratios increase more after the expansionary shock in the benchmark model than the model without prejudiced firms. This suggests more volatile procyclical wage and wealth responses for black workers than for white workers in the presence of $p$ firms. Moreover, consumption among both black and white workers responds more strongly in the benchmark model, leading to a more volatile consumption reaction following the expansionary TFP shock.

\subsubsection{Stochastic simulation}

We simulate the stochastic equilibrium of the economy over 10,000 periods under aggregate TFP shocks, and compare the outcomes with and without $p$ firms. While impulse responses capture the transition path following a one-time shock, stochastic equilibrium focuses on the economy's short-term responses to a series of aggregate shocks. The results are summarized in \Cref{tab: cyclegaps}.

The stochastic simulation results further support the insights from the impulse response analysis. We find that the presence of $p$ firms increases unemployment rate volatility for both black and white workers, but the effect is more pronounced for black workers. In the counterfactual economy without $p$ firms, both groups have an unemployment rate relative volatility of 0.18. With $p$ firms in the benchmark economy, the volatility for black workers rises to 0.26, compared to 0.22 for white workers. Similarly, the volatility of the black-white unemployment rate gap increases from 0.17 to 0.30, reflecting the disproportionate impact on black workers.

The wages of both black and white workers have dynamics similar to their unemployment rates. In particular, the presence of $p$ firms in the benchmark increases both the cyclicality and volatility of wages compared to the counterfactual without $p$ firms. However, this effect is more pronounced for black workers than for white workers. In our comparison of the benchmark and counterfactual economies, the inclusion of $p$ firms raises the cyclicality and volatility of black wages from 1.01 to 1.08, while white wages increase from 0.98 to 1. Overall, hiring discrimination not only heightens wage fluctuations but also widens the gap between black and white workers. These results support the findings of \citet{cajner2017racial}, \citet{boulware2019labor,boulware2024explains}, \citet{ganong2020wealth}, and \citet{ragusett2022re}, who argue that demographic differences alone cannot account for the racial disparities observed over the business cycle.

\begin{table}[htbp]
\centering
\caption{Impact of hiring discrimination over business cycle}
 \begin{tabular}{lcccc}
\toprule
& \multicolumn{2}{c}{Volatility} & \multicolumn{2}{c}{Cyclicality} \\
 & Benchmark & No $p$ firms & Benchmark & No $p$ firms \\ 
[0.5em] \hline\\
& \multicolumn{4}{c}{\textit{Unemployment rate}} \\
[0.5em]
Black & 0.26 & 0.18 & -0.69 & -0.70 \\
White & 0.22 & 0.18 & -0.69 & -0.70 \\
Black-White gap & 0.30 & 0.17 & -0.67 & -0.71 \\
[0.5em] \hline\\
& \multicolumn{4}{c}{\textit{Average wage}} \\
[0.5em] 
Black & 1.08 & 1.01 & 1.08 & 1.01 \\
White & 1.00 & 0.98 & 1.00 & 0.98 \\
[0.5em] \hline\\
& \multicolumn{4}{c}{\textit{Average wealth}} \\
[0.5em]
Black & 0.39 & 0.38 & 0.12 & 0.12 \\
White & 0.24 & 0.24 & 0.07 & 0.07 \\
Wealth ratio & 0.16 & 0.14 & 0.21 & 0.22 \\
[0.5em] \hline\\
& \multicolumn{4}{c}{\textit{Average consumption}} \\
[0.5em]
Black & 0.57 & 0.52 & 0.95 & 0.94 \\
White & 0.39 & 0.36 & 0.93 & 0.92 \\
Consumption ratio & 0.18 & 0.17 & 0.97 & 0.97 \\
\bottomrule
\end{tabular}

\vspace{3mm}
    
    \begin{minipage}{0.99\textwidth} 
{\footnotesize \textit{Note:} This table compares the responses of black-white unemployment rates, wages, consumption, and wealth gaps over the business cycle between the benchmark model and the model without $p$ firms.}
\end{minipage}
\label{tab: cyclegaps} 
\end{table}

We further examine the cyclical properties of wealth and consumption for black and white workers. Consistent with our earlier findings on labor market outcomes, the presence of $p$ firms increases the volatility of both the wealth and consumption ratios. Specifically, with $p$ firms, the volatility of the wealth ratio rises from 0.14 to 0.16, and that of the consumption ratio increases from 0.17 to 0.18. These results indicate that, beyond labor market variables, the volatility of racial gaps in other dimensions is also notably reduced when discriminatory hiring is eliminated.

In summary, our analysis shows that hiring discrimination amplifies racial disparities over the business cycle. The presence of $p$ firms increases the volatility of outcomes for black workers relative to white workers across multiple dimensions, including labor market variables, wealth, and consumption. These findings suggest that discriminatory hiring plays a key role in driving the cyclical differences in economic outcomes between racial groups.

\section{Conclusion}\label{sec: conclude} 

This study examines the effects of discrimination in the hiring of black workers on employment, wage disparities, and wealth accumulation in the US. We develop a search-and-matching model featuring firms with and without racial prejudices. Our findings reveal that racial discrimination in hiring significantly exacerbates wage, unemployment, and wealth gaps in the steady state and business cycles. In addition, discriminatory hiring places black workers disproportionately at the lower end of the wealth spectrum. Contrary to conventional discrimination theories, our analysis suggests that discriminatory hiring persists as an equilibrium outcome in frictional labor markets. Further, eliminating discriminatory hiring results in an increase in overall economic welfare. Our findings shed light on the enduring nature of black-white racial disparities in the US, offering insights into the interplay between discrimination, labor market dynamics, and wealth accumulation.

\clearpage

\newpage
\renewcommand{\thesection}{Appendix A} 
\def\theequation{\@A.\arabic{equation}}
\setcounter{equation}{0}
\setcounter{figure}{0}
\setcounter{table}{0}
\def\thefigure{\@A.\arabic{figure}}
\def\thetable{\@A.\arabic{table}}
\section{Stationary equilibrium}
\def\thesubsection{\@Appendix A.\arabic{subsection}}
\label{appA}

A stationary equilibrium consists of 
\begin{enumerate}
    \item Value functions of workers $\Big\{$$W_{np}(R,s,a)$, $W_{p}(wh,s,a)$, $U^I_{np}(R,s,a)$, $U^I_{p}(wh,s,a)$, $U^N(R,s,a)$$\Big\}$, and firms $\Big\{$$V_{np}$, $V_{p}$, $J_{np}(R,s,a)$, $J_{p}(wh,s,a)$$\Big\}$ 
    \item Corresponding asset policy functions of workers $\Big\{$$g_{np}(R,s,a)$, $g_{p}(wh,s,a)$, $g^I_{np}(R,s,a)$, $g^I_{p}(wh,s,a)$, $g^N(R,s,a)$$\Big\}$, along with the capital choice of producing firms $\Big\{$$k_{np}(R,s,a)$, $k_{p}(wh,s,a)$$\Big\}$ and vacancy choice of vacant firms $\Big\{$$v_{np}$, $v_{p}$$\Big\}$
    \item Wages $\Big\{$$\omega_{np}(R,s,a)$, $\omega_{p}(wh,s,a)$$\Big\}$
    \item Aggregate interest rate and labor market tightness $\Big\{$$r$, $\theta_{np}$, $\theta_{p}$$\Big\}$ 
    \item Labor income tax rate $\tau$ and unemployment insurance payout $\Big\{$$b_{np}(R,s,a)$, $b_{p}(wh,s,a)$$\Big\}$
    \item Dividends $d$
    \item Distribution over employment status ($e$), race ($R$), idiosyncratic productivity ($s$), and wealth ($a$), given by $\mu(e,R,s,a)$
\end{enumerate}
such that:
\begin{enumerate}
    \item $\Big\{$$W_{np}(R,s,a)$, $W_{p}(wh,s,a)$, $U^I_{np}(R,s,a)$, $U^I_{p}(wh,s,a)$, $U^N(R,s,a)$$\Big\}$ are the solutions to the worker's optimization problems (equations \ref{eqn:worker1}, \ref{eqn:worker2}, \ref{eqn:worker3}, \ref{eqn:worker4}, \ref{eqn:worker5}, \ref{eqn:worker6}, and \ref{eqn:worker7}), and $\Big\{$$g_{np}(R,s,a)$, $g_{p}(wh,s,a)$, $g^I_{np}(R,s,a)$, $g^I_{p}(wh,s,a)$, $g^N(R,s,a)$$\Big\}$ are the associated optimal decision rules for asset choice.
    \item $\Big\{$$J_{np}(R,s,a)$, $J_{p}(wh,s,a)$$\Big\}$ are the solutions to the producing firm's problems (equations \ref{eqn:firm3} and \ref{eqn:firm4}), and the corresponding capital choice is given by $\Big\{$$k_{np}(R,s,a)$, $k_{p}(wh,s,a)$$\Big\}$.
    \item Free entry of vacant firms, i.e., $V_{np} = 0$ and $V_{p} = 0$ determines the number of vacancies $\Big\{$$v_{np}$, $v_{p}$$\Big\}$, and hence labor market tightness $\Big\{$$\theta_{np}$, $\theta_{p}$$\Big\}$.
    \item Aggregate demand for capital equals aggregate supply, which in turn determines the interest rate $r$.
    \item Wages $\Big\{$$\omega_{np}(R,s,a)$, $\omega_{p}(wh,s,a)$$\Big\}$ are determined by Nash bargaining between the worker and the firm.
    \item Labor income tax rate $\tau$ solves to balance the government budget of the total unemployment insurance payout.
 
    \item Dividend $d$ is the total flow profits of producing firms, net of total posting costs of vacant firms.
    \begin{equation}
        d = -\kappa_pv_p -\kappa_{np}v_{np} + \int \mathrm{1}_{e=1,np}\text{   } j_{np}(R,s,a) d\mu + \int \mathrm{1}_{e=1,p}\text{   } j_{p}(wh,s,a) d\mu 
    \end{equation}
    where $j_{np}$ and $j_{p}$ refer to the flow profits of $np$ and $p$ firms respectively.
    \begin{equation}
        \begin{aligned}
            j_{np}(R,s,a) &= sf(k_{np}) - (r+\delta)k_{np} - \omega_{np} \\
            j_p(wh,s,a) &= sf(k_p) - (r+\delta)k_p - \omega_{p}
        \end{aligned}
    \end{equation}
    \item The distribution $\mu(e,R,s,a)$ is invariant and is consistent with the optimal decision rules of capital choice, the law of motion of idiosyncratic productivity, and the labor market flows.    
\end{enumerate}

\renewcommand{\thesection}{Appendix B} 
\def\thesubsection{\@Appendix B.\arabic{subsection}}
\def\theequation{\@B.\arabic{equation}}
\setcounter{equation}{0}
\setcounter{figure}{0}
\setcounter{table}{0}
\def\thefigure{\@B.\arabic{figure}}
\def\thetable{\@B.\arabic{table}}
\section{Comparing the impact of bargaining power and extreme wealth shock}
\def\thesubsection{\@Appendix B.\arabic{subsection}}
\label{appB}

Our model separately calibrates racial differences in bargaining power and the likelihood of extreme wealth destruction - features we refer to as non-market racial disparities. We conduct additional counterfactual analyses to assess the impact of these non-market factors relative to hiring discrimination. \Cref{tab: agg_nonmarket} compares the benchmark model to versions that eliminate non-market disparities by equalizing bargaining power and extreme wealth shocks across racial groups. When we set the bargaining power of black workers equal to that of white workers (Column 2, $\xi_{bl} = \xi_{wh}$), their bargained wages rise directly. However, the job-finding rate for black workers falls. This outcome occurs because non-prejudiced firms retain lower profits and, therefore, post fewer vacancies, which also reduces the job-finding rate and average wage for white workers. Consequently, the average wage gap shrinks to 20\% (a wage ratio of 80\%), accounting for roughly 20\% of the observed racial wage gap.

\begin{table}[htbp]
  \centering
    \caption{Aggregate impact of racial disparities from non-market factors}

\begin{tabular}{lccc}
\toprule
Moments & Benchmark & $\xi_{bl} =\xi_{wh}$ & $\epsilon_{bl}=\epsilon_{wh}$ \\
\midrule
\multicolumn{4}{c}{\emph{Households}} \\
job separation rate - black & 0.06 & 0.06 & 0.06 \\
job separation rate - white & 0.04 & 0.04 & 0.04 \\[0.5em]
job finding rate - black & 0.49 & 0.44 & 0.48 \\
job finding rate - white & 0.66 & 0.64 & 0.66 \\[0.5em]
unemp rate - black & 0.12 & 0.13 & 0.12 \\
unemp rate - white & 0.05 & 0.05 & 0.05 \\[0.5em]
mean wage - black & 1.59 & 1.70 & 1.65 \\
mean wage - white & 2.12 & 2.11 & 2.13 \\
mean wage ratio & 0.75 & 0.80 & 0.77 \\[0.5em]
mean wealth ratio & 0.29 & 0.30 & 0.77 \\
median wealth ratio & 0.33 & 0.33 & 0.88 \\
\midrule
\multicolumn{4}{c}{\emph{Firms}} \\
$p$ firm profit & 0.04 & 0.05 & 0.04 \\
\textit{np} firm profit & 0.02 & 0.02 & 0.02 \\[0.5em]
\textit{p} firm vacancy & 0.02 & 0.02 & 0.02 \\ 
\textit{np} firm vacancy & 0.05 & 0.04 & 0.04 \\[0.5em]
\textit{p} firm employment & 0.57 & 0.61 & 0.58 \\ 
\textit{np} firm employment & 0.37 & 0.33 & 0.35 \\
\midrule
\multicolumn{4}{c}{\emph{Labor Market}} \\
$p$ market tightness & 0.39 & 0.44 & 0.42 \\
$np$ market tightness & 0.71 & 0.60 & 0.67 \\
\midrule
\multicolumn{4}{c}{\emph{Aggregate Outcomes}} \\
Y & 3.07 & 3.06 & 3.08 \\
K/Y & 10.26 & 10.25 & 10.40 \\
average wage & 2.05 & 2.05 & 2.06 \\
unemp rate & 0.06 & 0.06 & 0.06 \\
\bottomrule
\end{tabular}

\vspace{3mm}

    \begin{minipage}{0.99\textwidth} 
{\footnotesize \textit{Note:} This table compares the benchmark steady state to a model with equal bargaining power $\xi_{bl} =\xi_{wh}$ and a model with equal extreme wealth shock $\epsilon_{bl}=\epsilon_{wh}$.}
\end{minipage}
\label{tab: agg_nonmarket}
\end{table}%

In Column 3 ($\epsilon_{bl}=\epsilon_{wh}$), we assign black workers the same extreme wealth shock as white workers, an external condition governing the accumulating wealth. The effect resembles assigning higher bargaining power to black workers. This is due to the importance of personal wealth in helping workers self-insurance against uncertain adverse outcomes \citep{nakajima2012business}. With a lower probability of losing their wealth, black workers can accumulate more personal wealth. Higher personal wealth gives black workers a higher reservation value when bargaining with firms. Effectively, black workers can negotiate for better wage outcomes (1.65 compared to 1.59 in the benchmark model). Similar to having higher bargaining power, the resulting job-finding rates decrease, though to a smaller extent. The lower extreme wealth destruction rate also translates to higher aggregate capital accumulation and increased aggregate output. It spills over to an increase in the average wage rate for all, raising the average black-white wage ratio to 77\%, thus explaining around 8\% of the racial wage gap. Since it directly raises the wealth position of black workers, the mean and median black-white wealth ratios increase drastically (77\% and 88\% from the benchmark levels of 29\% and 33\%).

\renewcommand{\thesection}{Appendix C}
\def\thesubsection{\@Appendix C.\arabic{subsection}}
\def\theequation{\@C.\arabic{equation}}
\setcounter{equation}{0}
\setcounter{figure}{0}
\setcounter{table}{0}
\def\thefigure{\@C.\arabic{figure}}
\def\thetable{\@C.\arabic{table}}
\section{Welfare calculation}
\def\thesubsection{\@Appendix C.\arabic{subsection}}
\label{appC} 

We follow \citet{krusell2010labour} in measuring the welfare effects of discriminatory hiring on black and white workers. We calculate the consumption equivalence in a counterfactual scenario without $p$ firms. Under the benchmark model, let $V(e,R,s,a) = \mathrm{E}_0 \sum_{t=0}^\infty \beta^t log(c_t)$ be the maximal value of the individual with employment status $e$, race $R$, productivity $s$, and asset $a$. In an alternative economy, let $\tilde{V}(e,R,s,a) = \mathrm{E}_0 \sum_{t=0}^\infty \beta^t log(\tilde{c}_t)$ be the maximal value of individuals with each corresponding state. We examine the welfare change between the two economies through consumption equivalence $\Omega$, following the equation:
$\mathrm{E}_0 \sum_{t=0}^\infty \beta^t log((1+\Omega)c_t) = \mathrm{E}_0 \sum_{t=0}^\infty \beta^t log(\tilde{c}_t)$. 

Under $\log$ utility, we derive $\Omega = \exp((\tilde{V}-V)(1-\beta))-1$. We aggregate the individual-level consumption equivalence, $\Omega$s, using the distribution of the counterfactual economy to calculate the average welfare change. We aggregate over the counterfactual distribution rather than the benchmark distribution because eliminating $p$ firms in the model removes the distribution of white workers from the states associated with $p$ firms. Aggregating using the benchmark distribution overstates the welfare change for white workers without accounting for this distribution shift.

\renewcommand{\thesection}{Appendix D} 
\def\thesubsection{\@Appendix D.\arabic{subsection}}
\def\theequation{\@D.\arabic{equation}}
\setcounter{equation}{0}
\setcounter{figure}{0}
\setcounter{table}{0}
\def\thefigure{\@D.\arabic{figure}}
\def\thetable{\@D.\arabic{table}}
\section{Measuring the business cycle properties for wage}
\def\thesubsection{\@Appendix D.\arabic{subsection}}
\label{appD} 

Aggregate data can be misleading when estimating wage cyclicality because shifts in worker composition over the business cycle tend to overweight lower-skilled (and lower-wage) workers during expansions, resulting in countercyclical bias. To overcome this issue, we use individual-level PSID data (1996-2014), which allows us to hold the composition of workers fixed over time.

Following the methodology of \citet{bils1985real}, \citet{solon1994measuring}, and \citet{devereux2001cyclicality}, we estimate wage cyclicality by regressing the change in individual log wages on the change in log GDP while controlling for key demographic characteristics. Our specification is as follows:

\begin{equation}
    \Delta \log w_{it} = \alpha + \beta \Delta \log Y_t + \gamma x_{it} + \nu_t + \upsilon_s + \epsilon_{it},
\end{equation}
where $Y_t$ is real GDP. The demographic variables $x_{it}$ include gender, education, labor market experience, and marital status. We also include year fixed effects, $\nu_t$, and state fixed effects, $\upsilon_s$. 

In this regression, the coefficient $\beta$ captures the cyclicality of wages. We then compute wage volatility as the standard deviation of wage growth after controlling for the demographics. The year fixed effects from the regression reflect aggregate wage growth and provide a measure of this volatility. This approach ensures that our estimates of wage cyclicality and volatility reflect true economic fluctuations rather than changes in workforce composition.

\renewcommand{\thesection}{Appendix E} 
\def\thesubsection{\@Appendix E.\arabic{subsection}}
\def\theequation{\@E.\arabic{equation}}
\setcounter{equation}{0}
\setcounter{figure}{0}
\setcounter{table}{0}
\def\thefigure{\@E.\arabic{figure}}
\def\thetable{\@E.\arabic{table}}
\section{Impulse responses}
\def\thesubsection{\@Appendix E.\arabic{subsection}}
\label{appE}

\begin{figure}[htbp]
\centering
\includegraphics[width=\textwidth]{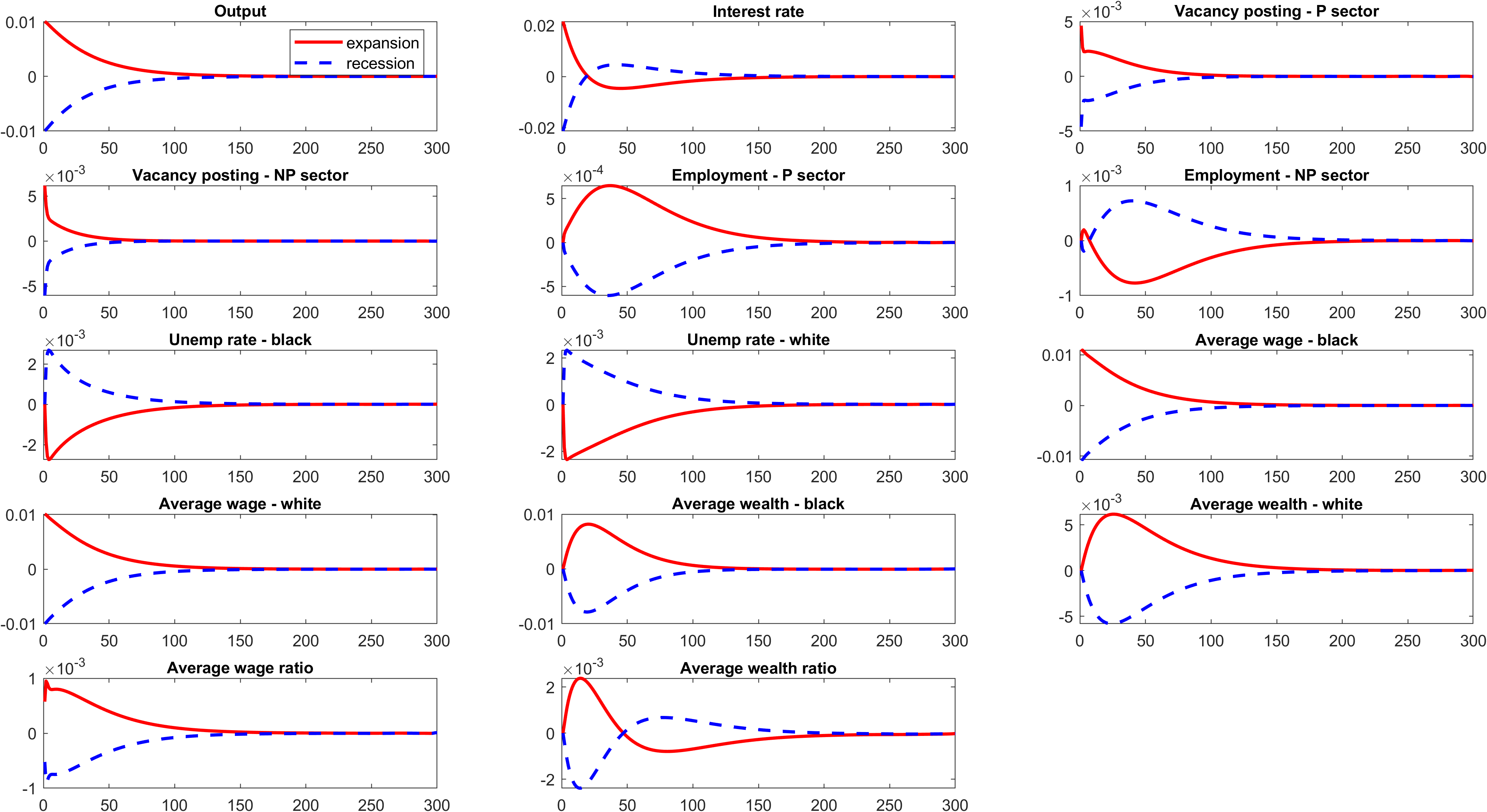}
\caption{Comparing impulse responses}
\vspace{3mm}

    \begin{minipage}{0.99\textwidth} 
{\footnotesize \textit{Note:} This figure plots the impulse response functions of the benchmark economy to a one percent increase and a one percent decrease in aggregate TFP at date zero. The red solid line denotes the responses to the expansionary shock, and the blue dashed line denotes the responses to the recessionary shock.}
\end{minipage}
\label{fig:irf_compare}
\end{figure}

We follow \citet{boppart2018exploiting} to simulate the business cycle moments of the model from the derivatives of the impulse response functions. For the impulse response functions to represent the numerical derivative, the magnitude of the MIT shock should be small. This method also requires the business cycle dynamics to be well approximated as a linear system. \Cref{fig:irf_compare} presents the impulse response functions for the benchmark model to a 1\% positive and a 1\% negative TFP shock at date 0. The shock gradually returns to a steady state with a persistence of 0.95. Our model produces symmetric impulse responses for positive and negative TFP shocks for all variables.

\clearpage

\bibliographystyle{chicago}
\bibliography{lib}

\end{document}